%% file: main.tex
\documentclass[journal,a4paper]{IEEEtran}
\usepackage{cite}
\usepackage{amsmath,amssymb,amsfonts}
\usepackage{graphicx,color,epsfig}
\usepackage{textcomp,comment}
\usepackage{multirow}
\usepackage{times,wasysym,bm,mathtools}
\usepackage{algorithmic}
\usepackage{psfrag,balance}
\usepackage{braket}
\usepackage{dsfont}

\title{\Huge Characterizing and Utilizing the Interplay between Quantum Technologies and Non-Terrestrial Networks
	
	\author{\IEEEauthorblockN{Hayder Al-Hraishawi\IEEEauthorrefmark{1},
			Junaid ur Rehman\IEEEauthorrefmark{1}, 
			Mohsen Razavi\IEEEauthorrefmark{2},
			and Symeon Chatzinotas\IEEEauthorrefmark{1}\\
			\IEEEauthorrefmark{1}\small Interdisciplinary Centre for Security, Reliability and Trust (SnT), University of Luxembourg, L-1855, Luxembourg.\\
			\IEEEauthorrefmark{2}\small School of Electronic and Electrical Engineering, University of Leeds, Leeds, LS2 9JT, U.K.}
	}
	\thanks{This work is financially supported by the Digital Europe Programme (DIGITAL) under the project of Lux4QCI - Luxembourg Experimental Network for Quantum Communication Infrastructure. For the purpose of open access, the authors have applied a Creative Commons Attribution 4.0 International (CC BY 4.0) license to any Author Accepted Manuscript version arising from this submission.}
	\thanks{Corresponding author: \textit{Hayder Al-Hraishawi (hayder.al-hraishawi@uni.lu)}.}
	}

\begin{document}
\bstctlcite{IEEEexample:BSTcontrol}
\maketitle

\begin{abstract}
Quantum technologies are increasingly recognized as groundbreaking advancements set to redefine the landscape of computing, communications, and sensing by leveraging quantum phenomena, like entanglement and teleportation.  Quantum technologies offer an interesting set of advantages such as unconditional security, large communications capacity, unparalleled computational speed, and ultra-precise sensing capabilities. However, their global deployment faces challenges related to communication ranges and geographical boundaries. Non-terrestrial networks (NTNs) have emerged as a potential remedy for these challenges through providing free-space quantum links to circumvent the exponential losses inherent in fiber optics. This paper delves into the dynamic interplay between quantum technologies and NTNs to unveil their synergistic potential. Specifically, we investigate their integration challenges and the potential solutions to foster a symbiotic convergence of quantum and NTN functionalities while identifying avenues for enhanced interoperability. This paper not only offers useful insights into the mutual advantages but also presents future research directions, aiming to inspire additional studies and advance this interdisciplinary collaboration.
\end{abstract}


\section{Introduction}\label{sec:introduction}
The recent breakthroughs in quantum technology development are paving the way towards establishing novel quantum networks based on quantum entanglement and superposition  phenomena. These networks aspire to enable numerous interconnected users in sharing information conveyed by quantum systems\cite{Pirandola2020}, and thus, establishing the envisioned concept of  quantum Internet (QI) \cite{Kimble2008, Qinternet_vision, Hosseinidehaj2019,Cacciapuoti2020,ILLIANO2022}. 
Quantum technologies, on a broader scale, hold the promise of introducing new development opportunities to classical communication systems including the potential for enhanced optimization techniques driven by advancements in quantum computing and the implementation of unconditionally secure cryptography, beyond the capabilities of classical systems \cite{DeLima2021}. 

However, a critical obstacle in deploying extended quantum networks over wide geographical areas is the exponential losses incurred by the fiber optic links due to light absorption and scattering\cite{Mehic2020}. In classical communications, optical amplifiers serve as repeaters to compensate such losses, but amplifying individual photons compromises their quantum information. 
Therefore, several research efforts have been directed toward the development of quantum repeater technologies to address this challenge. However, it is anticipated that these technologies will not be available on a large scale in the near future. Currently, the most practical near-term solution involves the use of free-space optical (FSO) links. In this direction, there have been recent explorations into incorporating non-terrestrial networks (NTNs) to extend the ranges of quantum communication. This is exemplified in \cite{ liao2017}, where a satellite link spanning 1200 km between two ground stations demonstrated a remarkable efficiency in terms of losses, surpassing an optical fiber link by 15 orders of magnitude \cite{YCL:17:Sci}. Thus, the integration of quantum technologies into NTNs, including aerial platforms and satellites, represents an interesting development step towards the establishment of  large-scale quantum networks \cite{simon2017}.

\subsection{Background}
Photons, the quantized units of light, stand as efficient  carriers of quantum information due to their unique characteristics and ease of manipulation. Their ability to travel at the speed of light ensures minimal latency in quantum communication systems \cite{levi2023}. 
The properties of indistinguishability and superposition inherent in photons make them well-suited for encoding and processing quantum states \cite{hochrainer2022}. Further, quantum entanglement, a nonclassical correlation between the quantum states of multiple photons, is beneficial in several quantum communication protocols \cite{cozzolino2019}.
Thus, leveraging quantum entanglement and teleportation establishes the foundation for prominent progress in quantum communications,  paving the way for a new era of information exchange \cite{Singh2021}.

 In the context of NTNs, the use of FSO links for connecting various NTN entities is favored in quantum communications due to the aforementioned reasons and the  unequivocal advantages of negligible background thermal radiation at optical frequencies \cite{trichili2020}. 
 Consequently, given the much lower channel losses and minimal decoherence in the space, NTN entities can be used as intermediate nodes for quantum communication between distant locations.  Additionally,
 it is important to highlight that quantum communication is also feasible in the terahertz (THz) band, albeit with its own set of challenges. These challenges include (i) a reduced transmission rate and (ii) the necessity for converters to enable the conversion between optical-THz and vice versa \cite{Ottaviani2020}.
 

Interestingly, the NTNs are increasingly recognized as a promising enabler in the evolution of communication systems, especially with the transition from fifth-generation (5G) to sixth-generation (6G) \cite{Giordani2021}. In the 5G era, NTNs contribute to overcoming terrestrial limitations, providing connectivity in remote or challenging environments and enhancing global coverage. Looking ahead to 6G, NTNs are expected to become even more integral, offering the potential for ubiquitous and resilient connectivity, high data rates, and support for emerging technologies like the Internet of Things (IoT) and distributed computing \cite{Iqbal2023}.
As research and development in NTNs and quantum technologies continue to advance, we can envision the emergence of transformative communication systems that improve our ability to connect and collaborate across the globe. NTNs, empowered by quantum capabilities, will lay the foundation for a new era of secure, high-speed, and resilient communication, shaping the future of our interconnected world. Thereby, the purpose of this work is providing an overview of key research progress in the direction of fostering a harmonious convergence among these diverse systems and technologies in order to fully harness this intriguing opportunity and shape a cohesive, system-oriented vision.


\subsection{Feasibility Studies}

In the context of NTN-based quantum communications, several studies and experimental demonstrations have explored the viability of leveraging satellites to extend the reach of quantum communication protocols on larger scales \cite{Villoresi2008,Vallone2015,Bourgoin2015,cheng2015,tang2016,Dequal2016,liao2017,Pugh2017e,Dequal2021,Chapman2022}. In particular, an experimental study of the conditions facilitating single-photon exchange between a satellite and a ground station has been presented in \cite{Villoresi2008}. The experiment has emulated a single photon source on a satellite and leveraged the telescope at the Matera Laser Ranging Observatory (MLRO) of the Italian Space Agency to capture the transmitted photons. Successful detections were achieved from the low-Earth-orbit (LEO) geodetic satellite Ajisai, with a perigee height of 1485 km. Additionally, the experiment in \cite{Vallone2015} has successfully demonstrated the polarization of a single photon over a satellite-to-ground channel can be maintained. The study realized quantum communication involving multiple satellites as quantum transmitters and the MLRO as the receiver, where the observed low quantum bit error ratio affirms the feasibility of implementing satellite-based quantum information protocols.

The feasibility of satellite-based quantum communication is demonstrated in \cite{Bourgoin2015} by implementing a quantum key distribution (QKD) protocol, involving optical transmission and complete post-processing, in the high-loss regime with minimized computing hardware at the receiver. By employing weak coherent pulses with decoy states, the production of secure key bits is demonstrated at up to 56.5 dB of photon loss. The feasibility of a satellite uplink is further illustrated in \cite{Bourgoin2015} by generating a secure key while experimentally emulating varying losses predicted for realistic LEO satellite passes at 600 km altitude. 
The design and implementation of an electronics platform for a quantum optics experiment on a  CubeSat to generate polarization-entangled photon pairs are presented in \cite{cheng2015}. This space-qualified platform marks an advancement in the conduct of quantum communication experiments with CubeSats.
The in-orbit operation of a photon-pair source aboard a 1.65-kg nanosatellite was reported \cite{tang2016}, demonstrating pair generation and polarization correlation under space conditions.
The study in \cite{Dequal2016} demonstrated the transmissions of single-photons from a medium Earth orbit (MEO) satellite to the MLRO ground station, with a link length exceeding 7000 km.
Achieving low quantum bit error rates using an active source on the MEO satellite and silicon detectors indicates the feasibility of implementing a QKD link.

More importantly, the deployment of the LEO satellite `Micius' for implementing decoy-state QKD was reported in in \cite{liao2017}, showcasing the accomplishment of a kilohertz key rate from the satellite to the ground over distances of up to 1,200 kilometers. This key rate surpasses expectations by around 20 orders of magnitude compared to using an optical fiber of the same length. 
Further, the study in \cite{Pugh2017e} successfully demonstrated QKD from a ground transmitter to an airborne receiver prototype. The developed pointing and tracking system exhibited the capability to establish and sustain an optical link with milli-degree precision over distances ranging from 3 to 10 km. During this operation, BB84 decoy-state signals were transmitted across the channel to the aircraft, which was in motion at angular speeds similar to those of a LEO satellite.

The Authors in \cite{Dequal2021} investigated the feasibility of secret key establishment in a satellite-to-ground downlink using continuous-variable encoding and utilizing standard telecommunication components certified for space environments, operating at high symbol rates. The obtained results showed positive secret key rates for LEO scenarios, while finite-size effects and high-loss channels may limit higher orbits. 
In \cite{Chapman2022},  entanglement-based QKD protocols are implemented, including the BBM92 model, achieving quantum bit-error rates  below 2\% in all bases. A low quantum bit-error rates execution of a higher-dimensional hyperentanglement-based QKD protocol was demonstrated, showing significantly increased secure key rates. This protocol is suitable for a space-to-ground link with Doppler-shift compensation as validated through rigorous finite-key analysis.

\subsection{Prior Related Works}
Over the last few years, several important surveys and review papers have explored the developments of quantum networks facilitated by satellites and aerial platforms, i.e., non-terrestrial communication systems \cite{Singh2021,Zhang2022,Kumar2022,Kumar2021,Sidhu2021,Hosseinidehaj2019,de2023,Kaltenbaek2021}. Specifically, the QI technologies, applications, and open challenges, along with the necessary infrastructure, have been  reviewed in \cite{Singh2021}, which also included some insights into satellite-based quantum communications.  In \cite{Zhang2022}, the authors provided a concise review and visionary outlook on quantum communication and networking, including the fundamentals, technological advancements, and challenges in quantum networks. It has also outlined a prototype for a quantum network architecture involving drones and satellites as extended repeaters. The study in \cite{Kumar2022} explored  the prospective use of drones in quantum networks, encompassing the design and feasibility of implementing quantum drone networks at various altitudes. Reference \cite{Kumar2021} surveyed the recent developments in quantum technologies and algorithms, and explored the role of quantum satellites in drone-based networks and communications, delving into the significance of quantum artificial intelligence and quantum machine learning for quantum networks.

The review paper in \cite{Sidhu2021} provided an overview of progress in space-based quantum technologies, offering a roadmap for the development of a global quantum network. It highlights the cost-effective coverage potential of small satellites, summarizes challenges in space quantum technologies, and discusses ongoing efforts to overcome these challenges.
The latest research developments pertaining to continuous-variable quantum communication through low Earth orbit (LEO) satellites have been discussed in \cite{Hosseinidehaj2019}, shedding light on various technical challenges in this domain.
The work in \cite{de2023} has outlined a high-level architecture for a generic quantum information networks (QINs), focusing on a satellite-based QIN to enable operational quantum communication services on a global scale to interconnect quantum computers, quantum sensors, and secured communication devices. 
The White Paper in \cite{Kaltenbaek2021} summarized the current state of quantum technology development with implications for space applications including the design, development, implementation, and utilization of quantum technology in the space sector.

\begin{figure*}[!t]
	\centering
	\def\svgwidth{450pt} 
	\fontsize{8}{4}\selectfont{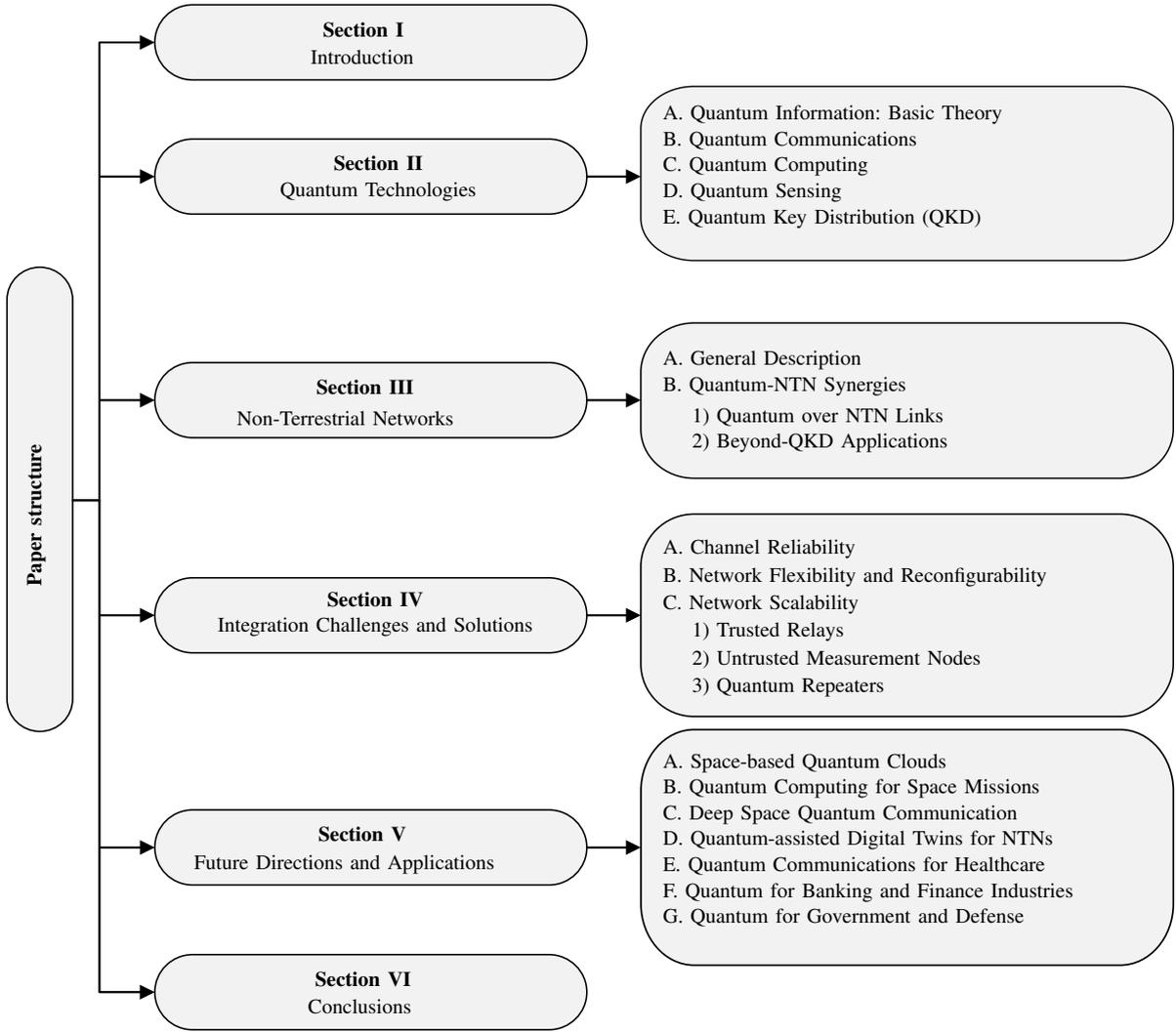}
	\caption{Structure and organization of the paper.}\label{fig:paper_structure}
\end{figure*}

\subsection{Scope and Contributions}
Although the studies mentioned above provide valuable insights, there are still crucial aspects in the integration of quantum technologies and NTNs that remain unexplored for advancing towards  \textit{reliable},  \textit{flexible}, and \textit{scalable} NTN-based quantum networks. For instance, the resemblance in equipment, such as optical links, required for interconnecting quantum nodes and NTN platforms presents untapped opportunities for further advancements in quantum NTN structures and functionalities. 
Thus, this article focuses on prominent quantum technologies, extending beyond the unconditional security aspects, and their deployment within versatile space-air-ground communication systems spanning different altitudes, layers, and orbits. We also emphasize on enhancing quantum-NTN interoperability, identifying and leveraging the mutual interplay between the evolving NTN architectures and quantum technologies. Through this, our paper contributes to a comprehensive understanding of these synergies, delving into essential features of quantum communication systems and their integration aspects with NTNs.  Hence, the key contributions of this paper can be outlined as follows:
\begin{itemize}
  \item Providing an up-to-date overview of the current knowledge and research works in the fields of quantum communication, NTNs, and their synergistic interactions. Presenting the quantum communication aspects and its diverse applications, along with NTN structures and interconnections, highlighting the opportunities that can be found in their convergence. 
  
  \item Delving into the symbiotic relationship between quantum technologies and NTNs, focusing on how shared equipment and technologies can foster seamless interoperability between the two domains.
    
  \item Exploring the critical integration hurdles and design aspects from the communication perspective, including \textit{channel reliability}, \textit{network flexibility}, and \textit{network scalability}, to effectively deploy NTN-based quantum systems, along with identifying potential solutions for each of these challenges.
    
  \item Presenting various promising research avenues and new applications that arise at the nexus of quantum technologies and NTNs.
   
\end{itemize}

This paper can serve as a valuable resource for understanding the current research contributions and the progress towards space-based quantum networks. It could potentially inspire further research endeavors in this field.

\subsection{Structure and Organization}
As presented in Fig. \ref{fig:paper_structure}, the remainder of this paper is organized as follows. In Section \ref{sec:Q_Comm}, quantum technologies are discussed. Section \ref{sec:Q_NTN} presents the key characteristics and architectures of NTNs in addition to the interconnecting synergies with the quantum devices. Next, the integration challenges along with its respective potential solutions are elaborated in Section \ref{sec:challenges}. Future research directions and opportunities are highlighted in Section \ref{sec:research_directions}. This article is finally concluded in Section \ref{sec:conclusions}.

\section{Quantum Technologies}\label{sec:Q_Comm}
Quantum technologies represent an advancing domain that leverages the fundamental principles of quantum mechanics to pioneer innovative tools and techniques with new capabilities. Within this field, several distinct advantages come to the forefront including the assurance of unconditional security, the potential for vast communications capacity, unparalleled computational speeds, and the ability to achieve ultra-precise sensing capabilities.  In this section, 
we  explore the fundamental concepts of quantum technologies, with a specific emphasis on those directly applicable to NTNs.


\subsection{Quantum Information: Basic Theory}
In classical systems, information is processed using bits, which are binary and can exist in states of either 0 or 1. Whereas,  quantum systems utilize quantum bits (qubits), which are the fundamental units of quantum information and can exist in multiple states simultaneously due to the principles of quantum superposition.
The state of an isolated quantum system is represented by a vector in a Hilbert space. A qubit is the simplest of such a quantum system, which is represented by \cite{Nielsen2010}
\begin{align}
	\ket{\psi} = \alpha \ket{0} + \beta\ket{1} = \begin{bmatrix} \alpha\\ \beta \end{bmatrix},
	\label{eq:qubit}
\end{align}
where $\alpha, \beta \in \mathds{C}$ with $\left| \alpha\right|^2 + \left| \beta\right|^2 = 1$ for normalization, $\ket{\cdot}$ is the Dirac notation for vectors, and $\ket{0} = \begin{bsmallmatrix} 1\\ 0 \end{bsmallmatrix}$ and $\ket{1} = \begin{bsmallmatrix} 0\\ 1 \end{bsmallmatrix}$ are the orthonormal vectors constituting the standard basis $\mathcal{B}_\mathrm{s} = \left\{ \ket{0}, \ket{1}\right\}$ of the two-dimensional vector space. Any qubit with $\alpha \neq 0, 1$ is said to be in the superposition of $\ket{0}$ and $\ket{1}$.

Once a quantum system is prepared in an arbitrary qubit state $\ket{\psi}$, it can be measured to extract \emph{no more than one bit} of classical information. In the simplest case, a measurement is performed by \emph{projecting} the state $\ket{\psi}$ onto the basis vectors of some orthonormal basis $\mathcal{B} = \left\{ \ket{\phi}, \ket{\phi^\perp}\right\}$, where $\ket{\phi^\perp}$ denotes the state orthonormal to $\ket{\phi}$. Upon measuring $\ket{\psi}$ of \eqref{eq:qubit}, the outcome will correspond to $\ket{\phi}$ with probability $\left| \braket{\phi \mid \psi}\right|^2$ or to $\ket{\phi^\perp}$ with probability $\left| \braket{\phi^\perp \mid \psi}\right|^2$, where $\bra{\phi} = \left( \ket{\phi}\right)^{\dagger}$ is the conjugate transpose of $\ket{\phi}$. It is easy to verify that measuring $\ket{\psi}$ in $\mathcal{B}_{\mathrm{s}}$ will give the outcome $\ket{0}$ with probability $\left| \alpha\right|^2$ and outcome $\ket{1}$ with probability $\left| \beta\right|^2$. More importantly, once measured, the state will no longer remain in the superposition of two states. Instead, it will assume the state corresponding to the obtained measurement outcome. In QKD, for instance, this collapse of state upon measurement can be used to detect the presence of an eavesdropper.

More interesting quantum phenomenon can be observed once we consider the state of multiple quantum systems. Entanglement is one such phenomenon, which can be observed in quantum systems consisting of as few as two qubits. The state of an arbitrary two-qubit system can be represented as
\begin{align}
	\ket{\psi}_{AB} &= \alpha \ket{0}_{A}\otimes \ket{0}_{B} + \beta \ket{0}_{A}\otimes \ket{1}_{B} + \nonumber \\ &~~~~~~~~~~~~~ \gamma \ket{1}_{A}\otimes \ket{0}_{B} + \delta \ket{1}_{A}\otimes \ket{1}_{B}\\
	&= \alpha \ket{00}_{AB} + \beta \ket{01}_{AB} + \gamma \ket{10}_{AB} + \delta \ket{11}_{AB},
	\label{eq:two_qubit}
\end{align}
where the state is normalized as before, subscripts indicate that the first qubit is part of system $A$ and second qubit is part of system $B$, $\otimes$ denotes the tensor (Kronecker) product, and $\ket{ij}$ is a shorthand notation for $\ket{i}\otimes \ket{j}$. 

Consider the two-qubit state of \eqref{eq:two_qubit} with $\alpha = \delta = 1/\sqrt{2}$ and $\beta = \gamma = 0$, i.e, 
$$
\ket{\phi^+}_{AB} = \frac{1}{\sqrt{2}}\left( \ket{00}_{AB} + \ket{11}_{AB}\right).
$$
The two qubits are correlated and in a joint superposition. Measuring one of the qubits will instantly define the state of the second qubit. This type of state is called an entangled state. Entanglement is a type of correlation that is known and experimentally verified to be stronger than any classical correlation \cite{Bel:64:PPF, PE:69:CHSH, HBD:15:Nat, AAA:18:Nat,Cacciapuoti2022}. It is a key ingredient in long-distance quantum communication and in several representative quantum communication protocols \cite{horodecki2009}. 

\subsection{Quantum Communication}
Quantum communication is a field that leverages the principles quantum mechanics  to enable  secure and efficient information transfer, ensuring transmissions that are unhackable and impervious to interception \cite{cariolaro2015}. 
The key enabling principles of quantum communication include quantum superposition, quantum entanglement, and the no-cloning theorem \cite{cozzolino2019}. 
Specifically, quantum superposition has potential to improve communication systems by enabling faster, more secure, and resilient data transmission. 
For example, the superposition property guarantees that any attempt to intercept the qubits will disturb their states, alerting the communicating parties to the presence of an eavesdropper \cite{Renner2023}. 
As discussed earlier, the quantum entanglement is a phenomenon where two or more particles become correlated in such a way that the state of one particle is directly related to the state of the other, regardless of the distance between them \cite{horodecki2009}. Entangled particles are used to establish a secret key between two parties, and any attempt to measure or disturb one entangled particle will affect its partner, providing a means to detect eavesdropping \cite{WIL:11:CUP}.

Quantum superposition can augment the channel capacity of communication systems by facilitating the simultaneous transmission of multiple bits of information. This property holds the potential to  enhance the data transmission capabilities of future communication networks \cite{Goswami2020}.
Furthermore, the concept of routing in classical networks involves determining the optimal path for data packets to travel from a source to a destination. In the context of quantum communications, quantum routing extends this concept to the transfer of quantum information. Quantum routing is a crucial element in the development of quantum communication networks, where qubits are transmitted instead of classical bits \cite{Caleffi2017}. In this, quantum routing utilizes the principles of quantum mechanics to optimize data routing in communication networks by identifying and selecting the most optimal paths for data transmission, reducing latency and improving network performance. Quantum superposition can be leveraged to explore multiple potential routes simultaneously, enhancing the efficiency of routing algorithms \cite{Shi2023}.

Quantum error correction is another crucial concept in the field of quantum communications, aiming to mitigate the impact of errors that naturally occur in quantum systems due to various factors such as noise and interference \cite{Lidar2013}. Unlike classical error correction, which involves duplicating information to detect and correct errors, quantum error correction relies on the principles of quantum mechanics to protect information. In quantum error correction methods, quantum states are encoded using a quantum code that introduces redundancy, enabling the detection and correction of errors without directly measuring the quantum states \cite{Yu2021}. This is achieved through the use of entangled states and quantum gates, which form a protective framework around the  information. Thus, quantum communications systems can maintain the reliability and fidelity of transmitted information by employing quantum error correction codes.

\subsection{Quantum Computing}
Quantum computing is a paradigm of computing that utilizes the principles of quantum mechanics to perform certain computations exponentially faster than classical computers, offering interesting advancements in solving complex problems in areas such as cryptography, optimization, and simulation \cite{De2020state}. It is important to note that while quantum computing shows great promise, practical, large-scale quantum computers are still in the early stages of development, and many technical challenges need to be addressed before widespread use becomes a reality. Researchers and industries are actively working on advancing quantum hardware, algorithms, and error correction techniques to unlock the full potential of quantum computing. 
Meanwhile, quantum devices that are currently accessible or expected in the near term are referred to as noisy intermediate-scale quantum (NISQ) devices \cite{Preskill2018quantumcomputingin}. These devices have proven to be more useful than initially anticipated, thanks to the advancement of specialized NISQ-era algorithms \cite{Bharti2022, ur2022variational,lee2022error}. Representative algorithms of NISQ-era computing are classical-quantum hybrid algorithms, such as variational quantum algorithms (VQAs) and quantum
approximate optimization algorithm (QAOA).


In recent years, researchers have focused on integrating quantum computing algorithms into classical communications, aiming to achieve certain performance goals like throughput, round-trip delay, and reliability at a lower computational complexity \cite{Botsinis2018}. Specifically, quantum computation offers distinct advantages in tackling problems requiring brute-force search in a large space by exploiting the essence of quantum mechanics. Several studies have highlighted the potential of employing quantum algorithms to tackle diverse optimization challenges, e.g., resource allocation \cite{Junaid2023} and system designs \cite{Rainjonneau2023}, presenting a more efficient approach \cite{Nielsen2010}. For instance, the binary knapsack optimization problem, that naturally appears when dealing with resource management in wireless communication systems, is formulated in \cite{Junaid2023} by utilizing a parallel QAOA framework, demonstrating a rapid identification of optimal solutions. Similarly, quantum algorithms are introduced in \cite{Rainjonneau2023}  for optimizing satellite mission planning problems, demonstrating their superiority over classical counterparts. 
Interestingly, the study in \cite{Córcoles2020} explored the opportunities of near-term quntum systems and indicated that existing quantum systems show substantial potential for effectively addressing certain dynamic and complex computing problems.

\subsection{Quantum Sensing}
Quantum sensing involves leveraging the principles of quantum mechanics to enhance the precision and sensitivity of sensors, allowing for more accurate measurements of physical quantities. Quantum sensors can exploit quantum properties such as superposition and entanglement to achieve unprecedented levels of precision in measuring various parameters such as time, frequency, magnetic fields, and gravitational forces \cite{Rademacher2020}. 
Particularly, quantum sensing is a	 promising technology within NTNs as quantum sensors can be deployed within satellite systems, offering high levels of sensitivity and precision for various applications such as remote sensing, navigation, and environmental monitoring. In this context, quantum sensors can measure different physical properties and much smaller quantities with a higher accuracy using miniaturized devices compared to the current sensors \cite{Degen2017}. 
Furthermore, quantum sensors can enhance the precision of atomic clocks, leading to improved timekeeping for satellite communication systems \cite{Wang2021}. Accurate time synchronization is vital for various applications \cite{Nande2023q}. Additionally, quantum sensors can provide ultra-precise navigation and timing signals, enabling NTNs to maintain accurate synchronization and positioning, crucial for seamless network operation and data communication \cite{Duan2021}.

Over the past decade, collaborative efforts among theorists, experimentalists, and engineers have explored the realization of functional NISQ sensing networks, yielding significant advancements in gravitational wave detection. The LIGO experiment, conducted in Michelson interferometers with up to 4 km arm lengths, incorporates environmental sensors to monitor disturbances \cite{Abbott2016}. Quantum shot-noise limit enhancements are achieved through the use of squeezed photons, boosting sensing resolution and detection efficiency. Other notable NISQ sensing networks focus on distributed quantum sensing, employing continuous and discrete variable quantum systems. In these experiments, inter-node and intra-node entangled states of up to 6 photons facilitate the estimation of spatially separated phase parameters with Heisenberg-limited precision \cite{Liu2021}. The utilization of quantum sensing networks is further exemplified in small-scale experiments, highlighting their applicability in verifying beneficial entanglement for synchronization within general quantum networks.

\subsection{Quantum Key Distribution (QKD)}
The QKD protocols utilize the principle of superposition of quantum states, collapse upon measurement, and the no-cloning theorem to distribute secret bits (keys) between spatially distant nodes \cite{BB:84:QC, FK:14:MC, TF:12:MT, CZW:22:ICST}. The main idea behind QKD comes from the fact that it is not possible to perfectly distinguish nonorthogonal quantum states \cite{ACM:07:PRL, CMM:08:PRA}\footnote{Two quantum states $\ket{\phi}$ and $\ket{\psi}$ are said to be orthogonal if $\braket{\phi \mid \psi} = 0$.}. The transmitter, Alice, can encode secret key bits in nonorthogonal quantum states and transmit over a quantum channel to the receiver, Bob. 
Upon successful reception, Bob measures the qubits in one of the predefined configurations. Next, Alice announces just enough information about the state preparation such that Bob can sift through the measurement configurations to decide which qubits were measured in the configuration compatible with the preparation. Finally, they compare a small subset of decoded key bits to estimate the error rate. Since an eavesdropper, Eve, cannot make perfect copies of transmitted qubits, it is not possible to replicate the measurements performed by Bob without introducing errors. This would enable Alice and Bob to bound the amount of leaked information to Eve based on the observed error rates. If such error rates are sufficiently low, Alice and Bob can use proper privacy amplification techniques to share a secret key among themselves. 

These nodes can utilize these secret keys to encrypt their messages to achieve secure communication. Since these keys are secret by the virtue of laws of physics and not by some computational complexity assumption, the achieved security is also called \textit{unconditional security} \cite{FK:14:MC, TF:12:MT}. 
In particular, once combined with well-known information-theoretically secure encryption methods, e.g., the one-time pad, the communicating parties can achieve information-theoretic security. Distribution of unconditionally secure key bits can be logistically challenging, especially for large amounts of data and over long distances. QKD addresses these challenges effectively.


Accordingly, during the past decades, QKD has received major attention from both research and industry communities where a remarkable progress has been made in experimental demonstrations. QKD has been realized with optical fibers and in FSO links using different degrees of freedom of photons including polarization, time-bin, energy and phase. Nevertheless, deploying QKD services worldwide is still a  highly intricate task due to the repeaterless PLOB bound \cite{Pirandola2017}.
As previously discussed, satellites and other NTN entities can be employed as trusted relays, leveraging their FSO links to facilitate key distribution over extended distances, and thus, provisioning a global QKD.


\begin{figure}[t!]\centering
	\includegraphics[width=0.5\textwidth]{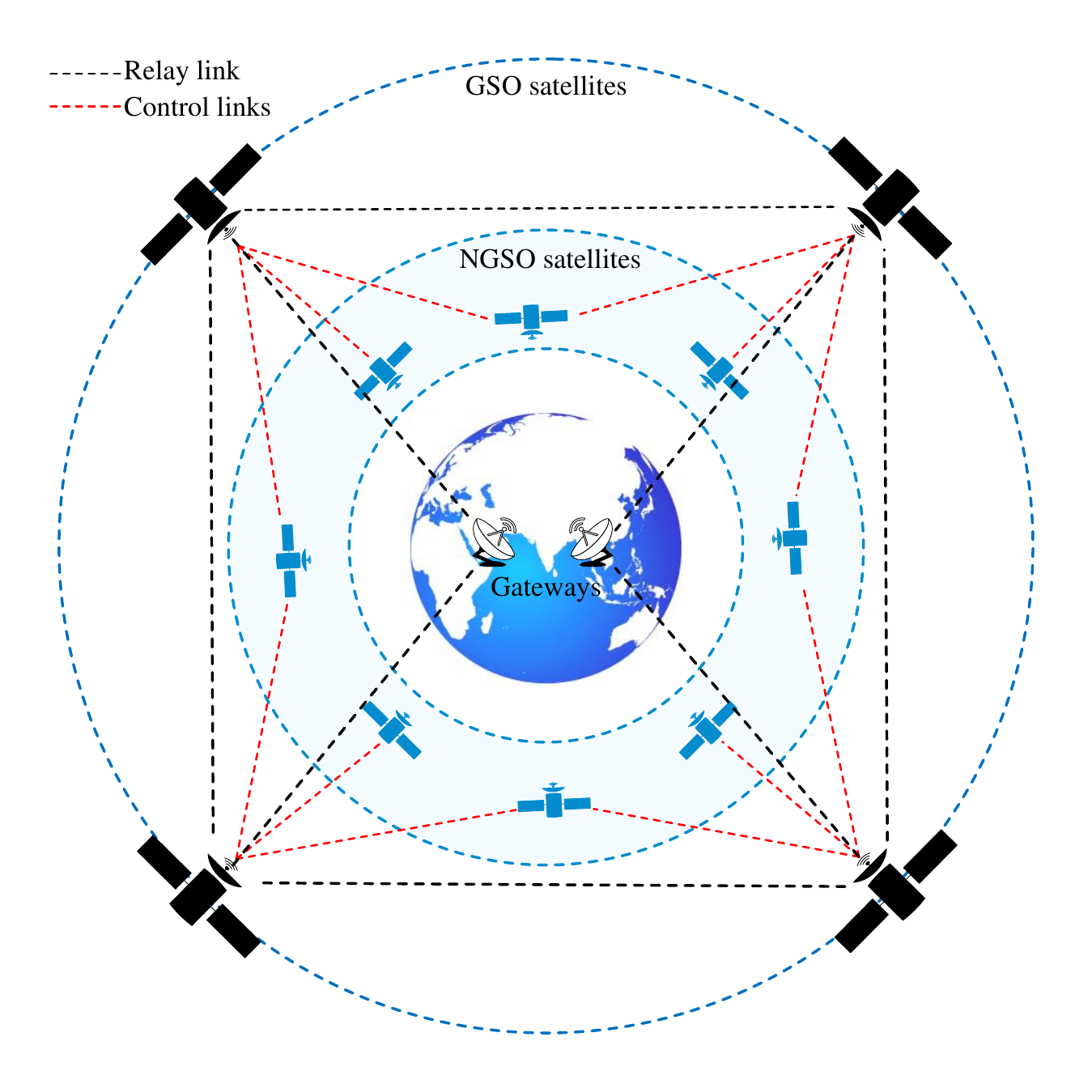}
	\caption{General schematic diagram of a multi-layered space-based network.}
	\label{fig:Q_SIN}
\end{figure}

\section{Non-Terrestrial Networks}\label{sec:Q_NTN}

\subsection{General Description}
Fundamentally, NTNs include various platforms that have different deployment options but they can be categorized based on their altitudes into two main categories: \textit{space-borne} and \textit{airborne}. The space-born platforms can also be classified based on their orbital geometry into  geostationary orbit (GSO) and non-geostationary orbit (NGSO) satellites, see Fig. \ref{fig:Q_SIN} for an illustration. GSOs are orbiting at the equatorial plane at an altitude of 35,678 km with an almost zero-inclination angle. Whereas, NGSO satellites on a geocentric orbit include the LEO, MEO, and highly elliptical orbit (HEO) satellites, which are orbiting constantly at lower altitudes than GSO satellites \cite{ITU2003}.  Airborne platforms, on the other hand,  involve unmanned aerial  vehicles (UAVs) that are placed at altitudes between 8 and 50 km, and high altitude platform systems (HAPS) that are deployed within 20 km altitude.

The recent and rapid growth of “NewSpace” industries makes the deployment of satellite mega-constellations feasible through reducing the costs of building, launching and operating small satellites, which significantly increases the number of satellites especially within the lower orbits \cite{al2021survey}. Similarly,
UAV and HAPS technologies have been growing in popularity and they are being developed and deployed at a very rapid pace around the world to offer fruitful business opportunities and new vertical markets \cite{Alladi2020}. This large number of diverse platforms imposes exceptional technical challenges on system control and operation, where they need to be built on an autonomous and dynamic network architecture \cite{Hayder2021}. Therefore, these various space and aerial platforms can be interconnected via inter-aerial links (IALs), inter-satellite link (ISLs) and inter-orbit links (IOLs) to construct a multi-layered integrated NTNs, which can support real-time communications, massive data transmission, and systematized information services \cite{Jung2023}.


Establishing multi-layered NTNs to connect multitude of platforms in different orbits/altitudes will enable combining multiple space/aerial assets to allow a more agile and efficient use of system resources \cite{Gupta2023}. 
This NTN architecture is more economically efficient and more suitable for delivering heterogeneous services and serving diversified applications. Furthermore, NTNs can satisfy the increasing complexity of application requirements with a minimum number of gateways on the ground \cite{Hassan2020}. 
For instance, utilizing the space-based Internet providers, such as Starlink and SES O3B, to provide broadband connectivity to the airborne and space-borne platforms can be a promising technique for nurturing the development of multi-layered NTN infrastructures \cite{SIN_2021}. 
Moreover, developing the seamless connectivity among multipurpose space-air-ground communications nodes over different altitudes, layers, and orbits will enhance the interoperability in future \cite{Michael2023}.

Nonetheless, the open connectivity and the interconnection complexity in such a dynamic architecture, as well as the lower computational capabilities of the small platforms, are seen as the most paramount hurdles in this development \cite{Wang2023}. Thus, quantum technologies can help overcome these challenges by providing enhanced security, increased computational efficiency, and novel communication paradigms, thereby paving the way for the seamless integration of quantum technologies with NTNs.

\subsection{Quantum-NTN Synergies}
This subsection presents the potential benefits and collaborative opportunities that lie at the intersection of these two rapidly evolving fields.

\subsubsection{Quantum over NTN Links}
\begin{figure}[t!]
	\centering
	\def\svgwidth{240pt} 
	\fontsize{8}{4}\selectfont{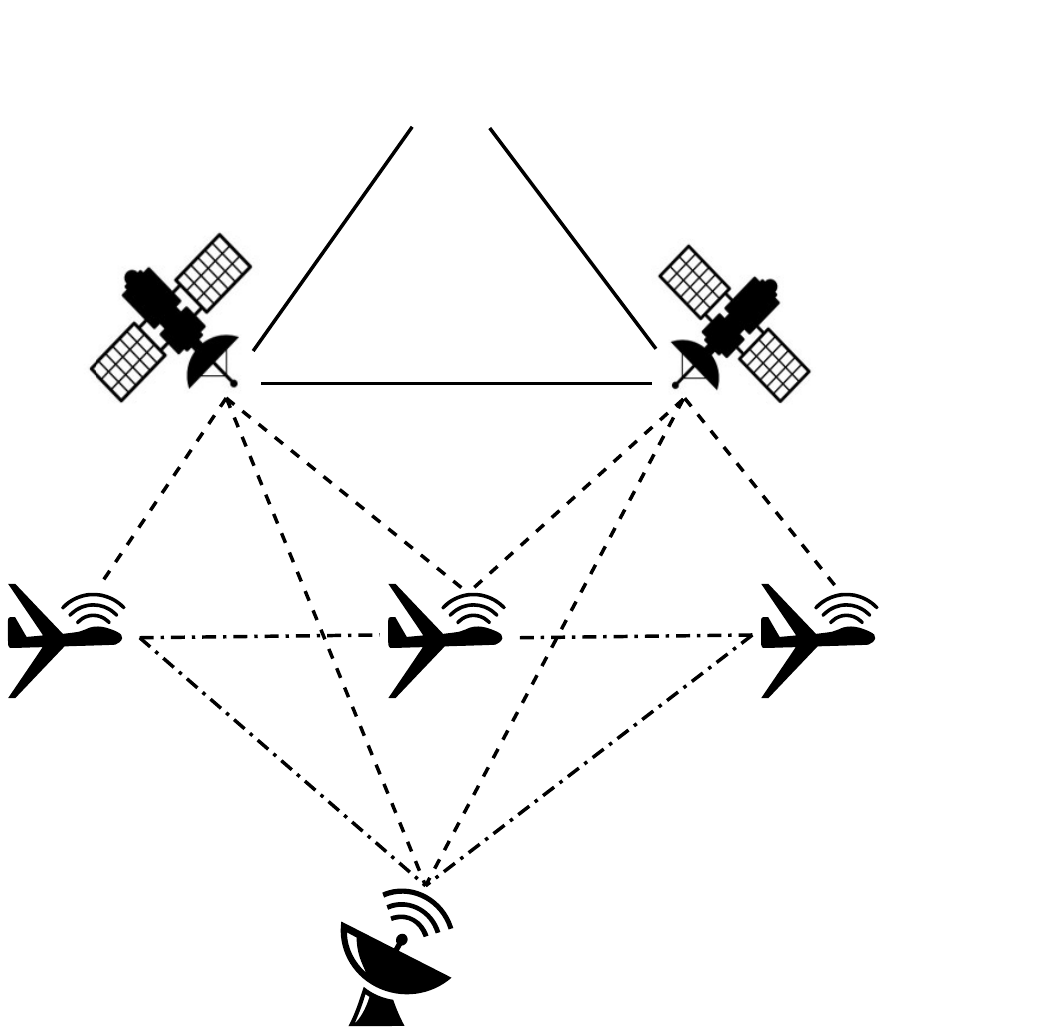}
	\caption{Multi-layered quantum NTNs.}
	\label{fig:Q_MLQN} 
\end{figure}

A single space-borne or airborne platform can connect two distant points with a maximum limit restricted by the platform altitude and the elevation angle through the atmosphere. Although GSO satellites have the ability to cover approximately a third of the globe, the achievable entanglement rates will be heavily deteriorated due to the vast communication range and low elevations at the extremities of the satellites trajectory, especially when considering dual path losses for non‐memory assisted quantum communication. Hence, global quantum connectivity can be realized through multi‐segment quantum links, which requires more complicated architectures such as entanglement swapping and quantum memories, {\it inter alia} \cite{bacsardi2018}. 
Thus, a multi-layered constellation of satellites and/or aerial platforms equipped with quantum devices (e.g. entanglement sources and quantum memories) can establish dynamically configurable multi‐link connections between any two points within the entire terrestrial and non-terrestrial integrated network, as depicted in Fig. \ref{fig:Q_MLQN}. Multi-layered quantum NTNs are essential for mitigating communication outages for sensitive equipment. The integration involves a network of satellites interconnected through quantum ISLs, providing extensive global coverage. Quantum communication tasks, e.g., QKD, are facilitated through FSO links . Additionally, encrypted classical links (ECLs) can be also enabled by quantum-secure classical communications within this comprehensive framework.


\begin{figure*}[t!]
	\centering
	\def\svgwidth{460pt} 
	\fontsize{9}{4}\selectfont{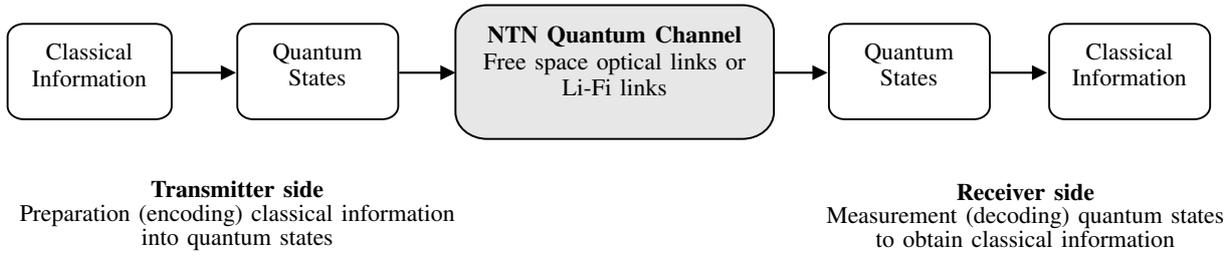}	
	\caption{Basic schematic diagram of an NTN quantum channel including encoding classical information into quantum states, Secure quantum transmission using free space optical or Li-Fi channel, and then, decoding the received quantum states to obtain the classical information.}
	\label{fig:Q_channel} 
\end{figure*}

Furthermore, the recent progress in quantum nonlinear optics, entangled photon generation, and single-photon detection has positioned NTNs as an important driver for the advancement of robust long-range quantum communication \cite{Yesharim2023}. In the context of quantum communication, the transfer of quantum states across spatial distances is achieved through quantum channels. These channels are in the optical domain and include optical fiber, FSO, or Li-Fi channels, as depicted in the schematic diagram in Fig. \ref{fig:Q_channel}. Within the framework of NTNs, next we will  review the offered features and connection schemes that will facilitate the seamless development of extended quantum networks.

\begin{itemize}
	\item \textbf{Optical Links}:
Optical communications technologies have an essential role in the multi-layered NTNs, especially within NGSO systems and mega-constellations, to establish efficient architectures using optical IALs, ISLs, IOLs, and ground-to-space/space-to-ground links. Furthermore, optical links can achieve higher data rates than conventional RF communications because the optical band provides much broader bandwidth, and thus, increases network capacity and alleviates the interference issues \cite{Gong2020}. Particularly, laser-based FSO ISLs and IOLs offer intrinsic high gains due to the narrow-beam nature of laser beams. Therefore, FSO technology is currently gaining momentum not only in experiments and demonstrations but also for commercial purposes in the context of connecting space missions. To react to this reality, the consultative committee for space data systems (CCSDS) has defined new specifications to deal with coding and synchronization of high photon efficiency links \cite{CCSDS1-19}. The objective of CCSDS is developing standards in wavelength, modulation, coding, interleaving, synchronization, and acquisition that are best suited for FSO communications systems \cite{CCSDS2-20}. Specifically, some working groups within CCSDS are dedicated on developing the coding and synchronization layer of a waveform supporting optical satellite-to-ground links along with optical modulation schemes to provide higher data rates up to 10 gigabits-per-second (Gbps) \cite{Edwards2022}.

Furthermore, FSO communications links have  been already experimented by the European Space Agency (ESA) and Japan Aerospace Exploration Agency (JAXA) for satellite-to-satellite link within the SILEX research program (Semiconductor Inter-Satellite Laser Experiment) \cite{tolker2002orbit}. In addition, NASA has recently launched the Laser Communications Relay Demonstration (LCRD) to showcase the unique capabilities of optical communications in space. In \cite{garcia2002preliminary, toyoshima2007results}, ground stations have been developed for optical space-to-ground links to investigate data transmission through the atmosphere. Whereas, an optical link between an aircraft and a GSO satellite was established and used to demonstrate a communication link in strongly turbulent and dynamic environment in \cite{cazaubiel2006lola}. All these studies and experimental demonstrations have  validated the feasibility of using FSO links to provide an unprecedented performance and the potential of being a favorable candidate for providing high-capacity connectivity to NTNs \cite{giordani2020}.

The evolution of FSO technology in NTNs can be further utilized by introducing quantum technologies for inter-orbit and intra-orbit connections as well as for downlinking to the gateways on Earth. From an implementation perspective, optical channels are typically used in quantum communication protocols owing to the negligible background thermal radiation at optical frequencies \cite{trichili2020}. Fortunately, both FSO connectivity and quantum communication share a symbiotic convergence in terms of the equipment needed for operation, and thus, they can each benefit from the technological developments in the other field. For instance, high-precision pointing systems needed for QKD applications can be used to improve FSO systems, while the adaptive optics modules developed for FSO will also improve the performance of quantum communication systems. Interestingly, there is also a considerable overlap in the design of high-rate encoders/modulators used in QKD and classical FSO systems, as both rely on on-off keying (OOK) and coherent optical modulation schemes. Thus, harnessing the synergetic interaction
between these two technologies can lay the foundations for novel communications paradigms based on an integrated terrestrial and NTN architectures and services.
\\

\item \textbf{Li-Fi Links}: Li-Fi technology is based on sending data using light waves as signal bearers with amplitude modulation of the light source. Li-Fi communications systems are able to utilize the vast optical spectrum to achieve peak data rates reaching the 10 Gbps level \cite{Sevincer2013}. Li-Fi extends the concept of visible light communications (VLC) to attain high speed bidirectional and fully networked wireless communications \cite{Spagnolo2020}. Moreover, Li-Fi systems offer more tangible benefits comparing to its RF counterpart such as affordable cost, low power operation, easy deployment, and point-to-point high data-rate communications, which provides high
bandwidth and operates in  license-free wide range optical spectrum. Additionally, Li-Fi communications can be used in RF-restricted areas such as hospitals, mines, and aircraft \cite{Wu2021}.
To this end, several feasibility studies have been conducted on using Li-Fi links within the satellite systems \cite{pavarangkoon2016,leba2017,Kalaivaanan2021}.

Furthermore, connecting the growing number of small-size, lightweight, low-power and low-cost satellites (e.g. CubeSats and nanosatellites) and aerial platforms in lower altitudes will be challenging due to the increased densification of NTNs \cite{Hayder2021}. Networking these different NTN entities requires highly survivable links capable of relaying and downlinking data in an efficient and plausible manner. In particular, it is mechanically challenging to deploy large parabolic antennas on small satellites equipped with RF radios in order to support high data rates. Additionally, the required pointing accuracy needed for laser communications presents a challenge to the form factor of CubeSats and nanosatellites due to the stringent size, mass, and power restrictions. Therefore, Li-Fi and VLC technologies can be seen as a potential solution to establish hybrid communications systems that are able to address these connection issues under certain circumstances \cite{Kalaivaanan2021}. Hence, Li-Fi in such scenarios  can surmount the NTN platforms’ limitations while avoiding the usual interference issues associated with RF systems \cite{Amanor2017}. 


The developments in quantum device technologies can reciprocally benefit Li-Fi and wireless optical systems. Within this context, multiple experimental and research studies explore the effect of different light-emitting diodes (LEDs) on Li-Fi performance. For example, a system based on OOK modulation that uses a white light LED, an analog pre-equalizer, and a post-equalizer is investigated in \cite{li2014550}, which achieves high-speed and low complexity VLC links. 
Similarly, multiple-quantum-well diode is studied in \cite{Jiang2017} to achieve on-chip optocoupling, which shows the capability of using these diodes for simultaneous emission and photo-detection in a full-duplex VLC system. In addition, the work in \cite{Wun2012} demonstrates the performance of a novel cyan LED, a light source for plastic optical fiber (POF) communications, in order to enhance the external quantum efficiency (EQE) and output power of this miniaturized high-speed LED, and thus, this setup improves the fiber coupling efficiency while achieving high data rates. In \cite{Le2021}, employing a QKD system for vehicular visible light communications (V2LC) networks is proposed by taking into account the VLC channel characteristics. The obtained results in these studies and experiments prove the feasibility of quantum Li-Fi integrated systems.

From the industrial communications networks, Light Rider company has recently unveiled quantum Li-Fi products that offer an unhackable  network connectivity. Further, VLC quantum fusion has been already considered for IoT applications to improve reliability and  security \cite{Suciu2020}. In this setup, quantum dots (QDs) are enabling materials for this integration owing to their easy customizable emission wavelength and superior quantum performance. Similarly, Li-Fi links along with the quantum technologies can be utilized in the emerging Internet of Space Things (IoST), which is a new class of small satellites used for data collection and equipped with limited onboard processing. Accordingly, empowering  near future small satellites and the various aerial platforms  with the Li-Fi networking capabilities along with embedding quantum technologies can offer further technical  improvements to establish wide-area quantum networks and to unlock new possibilities for secure and high-performance communication networks.


\end{itemize}

\begin{figure}[t!]
	\centering
	\def\svgwidth{190pt} 
	\fontsize{9}{4}\selectfont{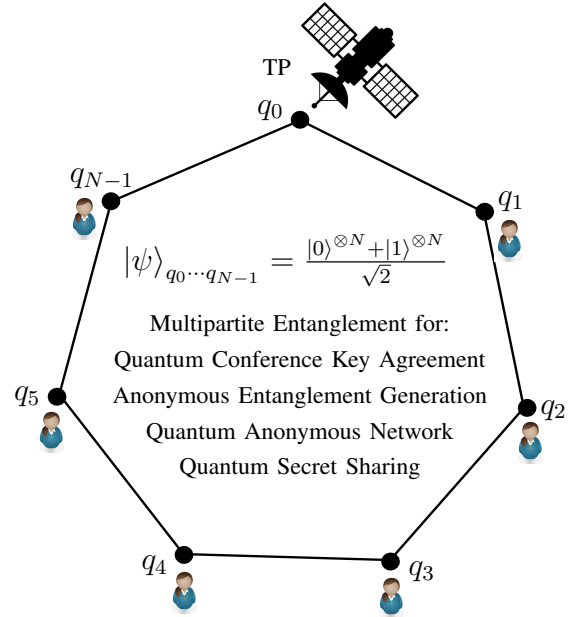}
	\caption{Quantum conference key agreement, quantum secret sharing, and quantum anonymous communications networks can be implemented in a similar network configuration with multipartite entanglement and NTN component acting as a central node.}
	\label{fig:Q_MPE} 
\end{figure}
\subsubsection{Beyond-QKD Applications}
\label{sec:beyondQKD}
The QKD is arguably the most mature research topic of utilizing quantum technologies in communications systems. However, there exist other use cases of quantum communication systems that provide unique advantages over their classical counterparts. In particular, some of these applications require the existence of a third-party (TP) with varying levels of trusts (trusted, semi-honest, untrusted). Such a requirement makes them interesting for implementation in an NTN framework where the non-terrestrial components can assume the role of the TP (see Fig.~\ref{fig:Q_MPE}). 
Subsequently, we present some captivating synergies arising from the integration of quantum NTNs in the directions of beyond-QKD use cases.

\begin{itemize}
	\item \textbf{Quantum Conference Key Agreement}:
QKD offers a bipartite protocol that allows two distant parties to securely establish secret keys. It is possible to utilize QKD to distribute same key among $N\geq 2$ participants by repeated applications of QKD. However, such a setup is resource inefficient, requiring many runs of the QKD protocol. However, the long-term vision of NTN quantum integrated network goes beyond mere bipartite links and includes various nodes.
Towards this direction, quantum mechanics allows the distribution of multipartite entanglement in a network, which acts as a common resource for the network participants. Quantum conference key agreement (QCKA) protocols typically leverage the multipartite entanglement to establish a common shared random key among $N \geq 2$ network participants in a single run of the protocol \cite{MGK:20:AQT}. QCKA allows the users to broadcast secure communications in a network. The rich structure of multipartite entangled quantum states opens the possibility for a wide variety of novel key distribution protocols within the NTN infrastructure, where the quantum correlations can be exploited to devise realistic multipartite schemes.


\item \textbf{Quantum Direct Secure Communications}
Quantum Secure Direct Communication (QSDC) is an interesting approach to secure communication using the principles of quantum mechanics without the need for a shared secret key \cite{Sheng2022}. The superposition phenomenon is exploited in QSDC to encode information and convey messages securely, and the entanglement plays a crucial role in QSDC to enable secure information transmission. Specifically, QSDC protocols need  to transmit photons in two rounds. In the initial round, two users share entanglement to establish the quantum channel. Subsequently, the sender of the message utilizes the dense coding approach to encode the message \cite{Bennett1992}. After encoding, one of the photons in each photon pair is sent back for Bell-state analysis (BSA) to read out the secret message. Thereby, QSDC protocols allow distant nodes to communicate directly in a secured fashion without requiring them to establish secret keys in advance and it does not require key storage \cite{BF:02:PRL, QRJS:17:PTEP, LL:20:NJP}. Hence, attacks on communication with employing QSDC obtain only random data without any useful information.
As an example, recent demonstrations, such as the integration of QSDC from a GSO satellite to a ground station \cite{Wang2021}, highlight its feasibility within NTNs. QSDC is continually improving security aspects and reinforcing the value propositions of quantum technology integration in NTN communication systems.

\item  \textbf{Quantum Secret Sharing}:
Quantum secret sharing is another interesting application for scalable architectures of quantum communication networks. In these schemes, a secret quantum state is shared among $N$ network participants in such a way that at least $k<N$ participants are required in order to reconstruct the state. In other words, quantum secret sharing splits a secret message of one user, called dealer, into several parts and distributes these parts among other users, called players, with each player receiving a part. The players can gain no information about the state if there are fewer than $k$ players willing for the reconstruction \cite{HB:99:PRA, CGL:99:PRL}.
Recent quantum secret sharing schemes allow assigning unequal weights $w_i$ to each party. Then, the secret can be unlocked if the sum of the weights of the parties willing to unlock the secret is greater than a predefined threshold $\omega$ \cite{CZC:21:SR}. 
Further, the work in \cite{Chengji2021} has analyzed the security and the performance of terahertz continuous-variable quantum secret sharing within ISLs, where the feasibility of a long-distance inter-satellite communications with multiple players has been proved. 
Accordingly, quantum secret sharing is an essential primitive for large-scale heterogeneous networks such as NTNs in order to secure multiparty communications

\item \textbf{Anonymous Quantum Communication}:
Additional interesting application of quantum communication is the provisioning of anonymity in networking tasks. Hiding the identity of communicating parties can be a desirable property in some scenarios. Entangled states have the unique property that local operations on any of the entangled particle can change the global state of the system regardless of the particle index. This change in the global state can be detected either by a global measurement or local measurements with partial announcement of results. This concept has been used to provide anonymity in several networking tasks including anonymous transmission of classical bits \cite{CW:05:AC}, qubits \cite{LMW:18:PRA, UMI:19:PRL}, anonymous ranking \cite{HWL:14:PRA}, anonymous notification \cite{KuS:21:QIP}, anonymous collision detection \cite{KKu:21:EQT}, and anonymous private information retrieval \cite{KKu:22:TCOM}. One prominent feature of these protocols is the guarantee of anonymity even when all network messages are monitored by an adversary and/or a malicious agent controls a major part of the network. 
Some of the aforementioned protocols have been demonstrated in laboratory conditions. As mentioned above, the requirement of a central node with varying levels of trust makes these protocols particularly suitable for deployment in NTNs. Recent demonstrations of satellite-based entanglement distribution paves the way for implementing these entanglement-based applications in the future \cite{YCL:17:Sci}.
\end{itemize}

\section{Quantum-NTN Integration: Challenges and Solutions}\label{sec:challenges}
Quantum communication over NTNs presents a viable solution to the challenges posed by large attenuation in long-distance fiber channels, opening the possibility of establishing intercontinental quantum networks. However, this architecture introduces a new set of challenges in terms of communication \textit{channel reliability}, \textit{network flexibility}, and \textit{network scalability}. Addressing these issues requires the development of innovative methods to enhance channel reliability and the formulation of new strategies for reconfigurable network management within multi-layered NTNs. In the upcoming subsections, we will explore the key technical challenges and design considerations for the efficient deployment of quantum NTNs, along with highlighting the respective potential solutions.

\begin{figure*}[t!]
	\centering
	\def\svgwidth{400pt} 
	\fontsize{10}{4}\selectfont{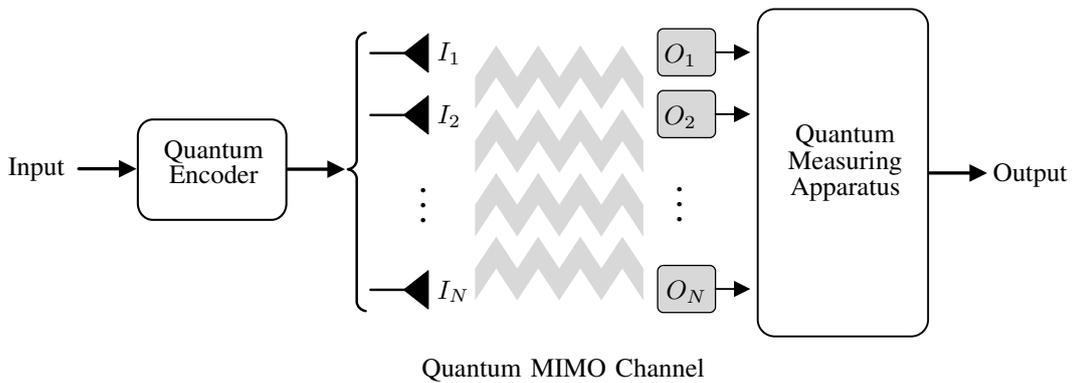}
	\caption{The schematic model of a quantum MIMO channel.}
	\label{fig:Q_MIMO}
\end{figure*}

\subsection{Channel Reliability }
FSO links are susceptible to atmospheric loss and scintillation effects caused by factors such as cloud blockage and variations in ionospheric electron density along the signal path. These environmental conditions contribute to fluctuations in signal intensity at the receiver \cite{kaushal2016}. It is important to note that FSO links are also vulnerable  to beam wandering and pointing errors, adding an additional layer of challenges to their reliability in adverse atmospheric conditions. Further, the movement of NTN units, and the constant requirement for pointing and tracking of these objects would also add to the overall channel loss. Another important factor in specifying the channel loss is the size and number of telescopes that can be used on a satellite, as well as on the ground stations. The randomness in all the above can also cause channel fading. In addition, the multipath time delay spread leads to time dispersion and frequency‐selective fading, whereas Doppler frequency spread leads to frequency dispersion and time‐selective fading. Moreover, the random effects of shadowing or diffraction from obstructions result in slow or large‐scale fading. 
in addition, satellite channels experience physical quantum noise due to time-variations of background radiations through different times of day and different seasons \cite{Pirandola2021,Rehman2023}.\\

\noindent 
\textbf{Potential solutions}\\
To enhance channel reliability, multicarrier transmission techniques and spatial diversity strategies, such as site diversity and multiple beam transmissions, can be applied in NTNs and FSO links \cite{erdogan2021}. Specifically, multiple-input multiple-output (MIMO) transmission techniques have drawn a significant attention in the satellite communications research due the offered high degrees of freedom \cite{Li2021,Hayder2023ICC}.

Additionally, MIMO transmission schemes have been explored in quantum optical wireless communication systems, referred to as quantum MIMO (q-MIMO) by the authors in \cite{Gabay2006} (illustrated in Fig.~\ref{fig:Q_MIMO}). It is important to note that while q-MIMO is one approach, there are alternative ways to apply MIMO techniques in quantum communication \cite{Kundu2021MIMODistribution}.
In contrast to the classical single-beam single-aperture configuration that is called SISO (single-input single-output), MIMO can realize spatial diversity by using a combination of multiple beams at the transmitter and multiple apertures at the receiver \cite{Almamori2017}. In the setting of \cite{Gabay2006}, MIMO communications is performed over quantum channels where classical information is transmitted through quantum states instead of classical electromagnetic field. The q-MIMO architecture is suitable for applications involving spatial diversity and optimal quantum digital receiver design, which will increase the reliability of quantum communication systems transmitting classical information over quantum channels. This architecture has promising aspects because it allows using positions of quantum antennas at the transmitter and quantum measurement operators at the receiver, which will allow for joined optimum fine-tuning of the overall system performance. 

MIMO techniques can also be used to deal with the fading nature of atmospheric channels \cite{Kadhim2016}.  For such channels, MIMO techniques are known to improve system performance. However, the advantages achieved  by using the MIMO concept come at the cost of utilizing more resources and increased system complexity. The MIMO techniques have not yet been fully investigated within the recent quantum advances and the multi-layered NTN structures, which can be exploited for reliable and secure high-speed communications. Thereby, it is essential to assess the feasibility and advantages of employing MIMO in the context of FSO communications and NTN systems. This could in turn result in performance improvement of both classical and quantum communication systems.

\subsection{Network Flexibility and Reconfigurability}

The seamless integration of quantum technologies into NTNs depends significantly on the flexibility and adaptability of the current network architectures \cite{Giordani2021}. Moreover, establishing a quantum link within NTN entities requires precision timing and time-tagging of the received photons, accurate pointing, robust filtering, and knowledge of the location, velocity, and range of both transmitter and receiver, which is an intricate task considering the dynamic propagation environment of NTNs. More importantly, the routing mechanism for coordinating quantum transmissions and classical communications is a crucial part for this integration, which should consider the unique features of both types of communications within this variable network’s topological structure. 

Furthermore, relaying of quantum keys among diverse network elements requires quantum devices to establish communication with the network control. This interaction is crucial for enabling flexible and efficient routing, guided by the quality-of-service (QoS) profiles of applications and the secret key rate generation of the corresponding links \cite{Hugues2019}.
Likewise, effectively managing resources across the quantum plane, data plane, and control plane poses a complex challenge that requires careful attention. Addressing this conundrum calls for the implementation of flexible and efficient resource allocation strategies. 
Particularly, safeguarding NTN communications with QKD mandates the presence of a quantum signal channel, a public interaction channel for secure key synchronization, and the conventional data channel \cite{Zhao2018}. The effective management of these three channel types necessitates the implementation of robust algorithmic solutions.\\

\begin{figure}[t!]\centering
	\includegraphics[width=0.5\textwidth]{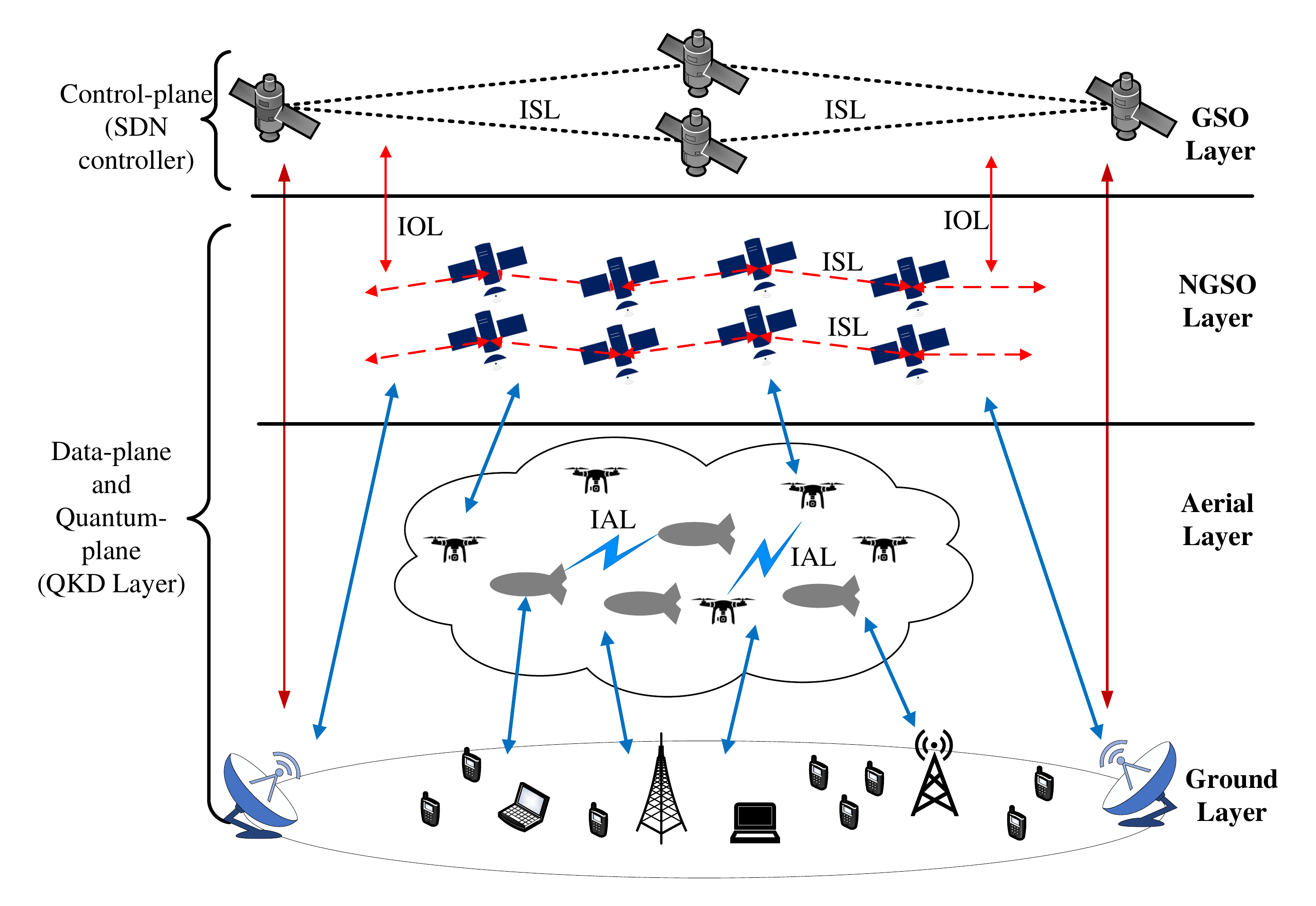}
	\caption{Quantum software-defined internetworking architecture in the terrestrial and non-terrestrial integrated systems.}
	\label{fig:Q_SDN_QKD2}
\end{figure}

\noindent 
\textbf{Potential solutions}\\
In this context, software defined networking (SDN) is a well-known paradigm for enabling flexible and programmable network configuration in order to improve system performance, management, and monitoring \cite{xu2018software}. SDN enables agile and efficient network services through innovative and advanced resource management techniques. Within NTN context, SDN can play an important role owing to the offered operational flexibility, scalability, and the end-to-end service provisioning \cite{SIN_2021}. Furthermore, embedding SDN into satellites and aerial platforms can facilitate the interaction between non-terrestrial and terrestrial wireless networks, which allows addressing several coexisting challenges. Additionally, SDN paradigm seems to be more suitable to deal with the complexity and dynamicity of multi-layered NTNs, where SDN controllers can be distributed over higher orbits  to simultaneously manage both the classical and quantum parts of the network within the active space and aerial nodes in lower altitudes as shown in Fig.~\ref{fig:Q_SDN_QKD2}. For instance, the SDN architecture can be designed to allow a controller to centrally orchestrate the quantum resources for optimizing the key allocation and systematizing the establishments of a direct channel or multi-hop links based on demands, visibility, and channel conditions.


The SDN technology can help with the management of different tasks we need to control in quantum communication systems. This includes key exchnage mechanisms, entanglement generation/distribution schemes, and efficient swapping procedures. In all cases, embedding SDN technologies into quantum networks would allow for provisioning of accurate control and management. Several research works in the open literature have shown that SDN integration is beneficial to QKD networks by customizing network configuration and designing efficient routing protocols \cite{Picchi2020}. Specifically, SDN can provide a constant instantaneous monitoring of the quantum parameters such as quantum bit error rate and secret key rate, and flexible configuration of optical paths to ensure the continuous distribution of quantum keys in the network. 
Furthermore, SDN allows the deployment of advanced resource allocation and control algorithms for load balancing, network slicing, and quantum-aware path computation, regardless of the underlying infrastructure.
Thus, introducing the SDN model is a mutually beneficial arrangement that opens the road to a seamless convergence between NTNs and quantum technologies.

\subsection{Network Scalability}
While NTNs present promising solutions for scaling up quantum communication networks, significant technical challenges persist, hindering the full realization of these integration benefits. Specifically, FSO and Li-Fi links transmitting to the ground rely on atmospheric channels as a propagating medium whose properties are random functions of space and time. 
This renders the quantum communication to a random process depending on weather and geographical locations \cite{deng2017}.
Moreover, various unpredictable environmental factors, including clouds, snow, fog, rain, and haze, among others, can lead to signal attenuation and reduce communication distances. However, it is worth noting that space-based communication links offer distinct advantages, characterized by negligible propagation losses as signals traverse the vacuum environment \cite{LCH:18:PRL}. This distinction underscores the potential resilience and efficiency of space-based quantum communication systems, especially when contrasted with the susceptibility of terrestrial links to atmospheric conditions.
Specifically, ISLs and IOLs are immune to weather conditions as satellite orbits are usually situated far above the atmosphere. However, the primary challenge arises from ensuring link availability when satellites are in motion with varying relative velocities.
As a result, the inherent transmission losses in actual optical channels, coupled with the difficulties in inter-satellite and inter-orbit communications, have the potential to diminish or entirely disrupt quantum entanglement across remote nodes. This jeopardizes the scalability promised by NTNs in achieving long-range quantum networking.\\

\noindent 
\textbf{Potential solutions}\\
To surmount these limitations, the NTN infrastructure can leverage various quantum devices and technologies. In the following, we will explore the most promising approaches aimed at achieving foreseeable scalability and ensuring a cohesive hybrid deployment landscape.
\subsubsection{Trusted Relays}
\begin{figure}[t!]\centering
	\def\svgwidth{225pt} 
	\fontsize{8}{4}\selectfont{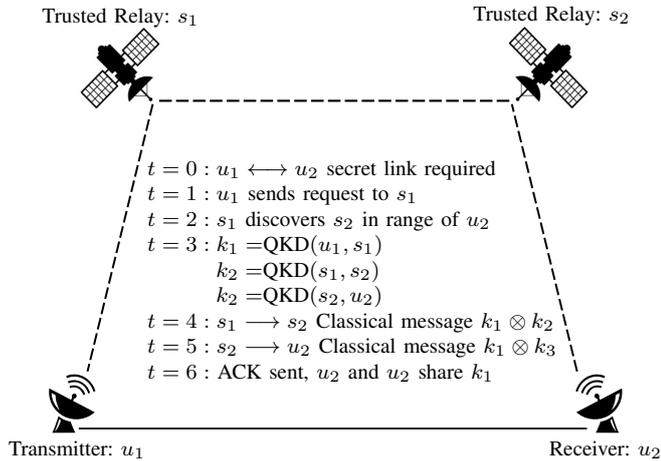}\vspace{4mm}
	\caption{Trusted-relay based quantum NTNs can extend the distance of QKD beyond the reach of ordinary links. Here an example scenarios is depicted where two satellites $s_1$ and $s_2$ assist two distant ground stations $u_1$ and $u_2$ to establish secret keys.}
	\label{fig:Q_QKD}
\end{figure}

Reliable and efficient transmission of quantum information at global distances is a daunting task. However, for some specific applications of quantum communication, it is possible to somehow decode-reencode the (classical) message encoded in the quantum states at trusted relays strategically positioned between communicating nodes. QKD is one example of such applications \cite{Mehic2019}. For establishing a key between Alice and Bob, with a trusted relay, Charlie, one can use two runs of QKD. Charlie establishes two random keys $k_1$ and $k_2$ with Alice and Bob by QKD, respectively. Then, Charlie sends the bit-wise exclusive-OR of the two keys to either of the two communicating parties, who can now obtain the key established with the other party. This key can be used for subsequent secret communication between Alice and Bob. Fig.~\ref{fig:Q_QKD} shows an example scenario with two satellites and two ground users. As it is clear from the above procedure, this approach is completely scalable with two serious drawbacks: 1) this approach is not generally applicable to all quantum communication protocols, and 2) very high level of trust is assumed, which may be hard to justify in general. However, this is one of the currently feasible solutions, which has been demonstrated with a LEO satellite assuming the role of a trusted relay and establishing secure keys between two ground stations 7600 km apart \cite{LCH:18:PRL}.

\subsubsection{Untrusted Measurement Nodes}
Measurement device independent (MDI) is an important framework to constitute a quantum network with an untrusted network server/relay, which can provide an enhanced security performance compared to traditional QKD \cite{Xu2015}. Thereby, even with untrusted NTN platforms, quantum security can still be guaranteed in some cases using the MDI concept. The main idea in MDI QKD is to design the communications protocol in such a way that no assumption on the trustworthiness of measurement devices is required. Both communications parties prepare and send their signals to the untrusted measurement device, which announces the measurement outcomes. The protocol is designed in such a way that the announcement of incorrect measurement results would show itself in the observed error rates, indicating malice or malfunction. Whereas, the announcement of correct measurement results does not pass any information to Eve/untrusted node about the actual key bits being exchanged between the two parties. 

The implementation of MDI QKD in the NTN setting is not without its own challenges. First, it requires us to use the uplink configuration, which is known to be more lossy than the downlink one, in satellite-to-ground settings. It also requires to have the measurement devices on board the satellite, which are often more complicated systems than the source in the MDI-QKD protocol. Synchronization is also a challenge given that the photons sent by the two users have to reach the satellite at the same time. That said, implementation of MDI QKD via a satellite node is a prelude to the implementation of memory-assisted QKD \cite{Panayi_2014, Gundogan2021} and quantum repeaters in space, and can therefore be part of the global efforts to make quantum communication services accessible worldwide.

\subsubsection{Quantum Repeaters}
The fundamental solution to the issue of scalability is utilizing quantum repeaters, which are the essential parts of future quantum communication systems \cite{mastriani2020,ghalaii2020}. In the case of QKD, quantum repeaters enable end-to-end security for QKD users. Conventional quantum repeaters aim at creating entanglement within smaller segments, followed by entanglement swapping (ES) at intermediate nodes to extend the entanglement to longer distances \cite{razavi2009}.
Embedding quantum repeaters in NTNs allows for realizing entanglement distribution over large distances with a smaller number of intermediate nodes as compared to terrestrial communications systems \cite{liorni2021}. In principle, quantum repeater nodes can be placed on board a satellite or an aerial platform, with photonic channels enabling entanglement distribution among orbiting/flying nodes. ES procedures can then be done at such nodes \cite{Pirandola2020}. Creating an end-to-end entangled state, when the nodes are moving in space, would inevitably add an extra layer of complexity to the design of the NTN based system. Therefore, it is critical to efficiently optimize the quantum repeater schemes, which is nontrivial because the number of possible schemes that can be performed grows exponentially with the number of links or nodes \cite{Goodenough2021}.

Quantum repeaters enable the implementation of all quantum networking tasks that require preshared entanglement between distant nodes as an important prerequisite; see Sec.~\ref{sec:beyondQKD} for some examples. This makes the ability of quantum nodes, i.e. end nodes and repeaters alike, to store and efficiently utilize quantum entanglement a crucial functionality. More importantly, tasks such as quantum teleportation and entanglement swapping are elementary and fundamental in nature for the basic working of a quantum communication network \cite{Cacciapuoti2020}. To fulfil the requirement of preshared entanglement, quantum memories that are capable of storing quantum information from generation to utilization while maintaining acceptable fidelity levels are required. 
A quantum memory is a device that can store an incoming photon and efficiently retrieve the same photonic state on-demand without disturbing the quantum state. Thus, an NTN-deployable quantum memory would be essential for long-range quantum communication and for performing QKD across global distances without intermediate trusted nodes. 

The development and deployment of quantum memories, by itself, is a huge technical challenge even for terrestrial applications. One of the key issues is the required coherence time of quantum memories. The authors in \cite{RPL:09:PRA} consider a quantum repeater network with a large number $(\gg 1)$ of quantum memories at each node to minimize the waiting time due to classical communications. Their optimistic estimates indicate that coherence times in the excess of 10~ms are sufficient for a 1000~km fiber-based repeater network. Meanwhile, experimental demonstrations have been performed for quantum memories of coherence times well above this limit, e.g., from 1.3~s to six hours \cite{ZHA:15:Nat, Ran:18:NP, ZHA:15:Nat}. However, these quantum memories typically store quantum information in matter qubits and/or require cryogenic temperatures. Storage in matter qubits requires development of efficient interfaces between flying and matter qubits, which in itself is a challenge \cite{ABB:21:PRXQ}. Requirement of cryogenic temperatures make utilization of quantum memories a challenge in non-laboratory conditions. Availability of practical quantum memories with sufficient coherence times will enable not only the long-distance quantum communication but also will greatly diversify the suite of useful quantum communication protocols that can be implemented in such networks.\\

There are multiple competing approaches that are being considered as candidates for quantum memories with different strengths and weaknesses \cite{ZHA:15:Nat, WUZ:17:NP}. In this direction, a variety of different dopant/host combinations  have been studied for various quantum mechanical phenomena, and many elements necessary for a practical quantum memory have been shown, such as long storage times and high efficiency optical storage and recall. However, the research still focuses on optimizing single parameters, while a system demonstrating all necessary aspects simultaneously remains to be developed. To this end, there are ongoing efforts for the integration of these memories in quantum networks such that the modality of quantum memory can be made independent of the operating modality of the quantum network \cite{ABK:21:PRXQ}. Once developed, these quantum interconnects will allow seemless interface between quantum nodes working with different modalities. The experimental developments can be boosted by the improved funding opportunities and allocation of more funds targeting the key components of the quantum network architecture. Meanwhile, theoretical efforts can be concentrated to the development of useful quantum technologies realizable with currently available hardware. Prepare-and-measure type protocols, e.g., BB84 are well within the reach of current experimental capabilities. Developments of novel quantum applications with the same structure can provide a boost in the utilization of quantum technologies in near future.

The above developments could, however, take years to be space ready. Another challenging task in the adoption of quantum communication networks in NTNs is then the limited number of demonstrable network tasks of practical interest. Most applications we mentioned in earlier sections,  e.g., quantum secret sharing, as well as other emerging ones such as quantum secret comparison, quantum oblivious transfer, and quantum voting,  require quantum resources beyond current technological reach, e.g., large amounts of long-term entanglement and error-corrected communication and storage \cite{Zhang2019}. These quantum resources are not likely to be available very soon \cite{Preskill2018quantumcomputingin}. In the mean time, it is essential to develop quantum network applications that are less resource-demanding and can be demonstrated with the currently available or near-term quantum communications equipment.

\section{Future Directions and Applications} \label{sec:research_directions}
The disruptive potential of the convergence between quantum technologies and NTNs extends beyond secure communications, offering new horizons for digital innovation. This section will explore various promising research directions and novel applications that emerge at the intersection of quantum technologies and NTNs.

\subsection{Space-based Quantum Clouds}
The concept of space-based clouds envisions satellites not only as relay devices but also as platforms for establishing data storage and processing paradigms in GSO and NGSO. The main advantage of space-based data centres is the absolute immunity against natural disasters on Earth  \cite{Jia2017}. In the context of multi-layered NTNs, geographical boundaries and terrestrial obstacles pose no hindrance to global data transfer. Mega-corporations with intercontinental sites can leverage space-based clouds to share massive data, benefiting from faster transmissions compared to terrestrial networks. 
Beyond this, incorporating quantum technologies enhances security and provides quantum computing capabilities for big data applications. This ecosystem creates an accessible environment for quantum algorithm development worldwide, potentially fostering emerging quantum-as-a-service providers \cite{Raddo2019}. Additionally, given the extreme high costs of hosting and building quantum computing services, space-based quantum clouds can improve the financial viability through allowing simultaneous access for multiple beneficiaries and users, and hence, increase machine utilization.

\subsection{Quantum Computing for Space Missions}
One of the major challenges faced by CubeSats and small satellites in lower altitudes is the limited processing capabilities of their onboard processors \cite{lovelly2014}. This constraint hampers the execution of complex processing tasks, including online optimization of resource allocation, Earth observation data processing, and IoT data aggregation. To address this, quantum technologies, coupled with space-based quantum clouds, offer a viable solution by offloading the computational burden from small satellites. Establishing a space quantum network interconnected via FSO links leverages the advantages of FSO over RF systems, coupled with the exceptional computational capacity of quantum servers, ensuring enhanced security. This setup not only mitigates latency issues, particularly for resource-intensive and time-sensitive applications but also positions small satellites as space-based quantum sensors. These quantum sensors on satellite nodes significantly enhance the practical performance of navigation and Earth observation systems \cite{Muller2020}. Specifically, Earth observation missions can measure small-scale variations in Earth's gravitational field, providing valuable insights into phenomena such as water flows, ice movement, and continental drifts.

\subsection{Deep Space Quantum Communication}
 Deep space quantum communication (DSQC) \cite{Mohageg2022} represents a pioneering frontier in the integration of quantum technologies with NTNs. As humanity ventures into deep space exploration, conventional communication methods face challenges such as signal degradation, propagation delay, and limited bandwidth. DSQC leverages quantum principles to address these challenges, promising secure and efficient communication over vast cosmic distances. The unique features of quantum entanglement and superposition can potentially enhance the reliability and speed of information transfer in the extreme conditions of deep space. Integrating DSQC within NTNs can revolutionize interplanetary communication, providing unprecedented capabilities for transmitting sensitive data and enabling novel applications in deep space missions. The exploration of DSQC within the context of NTNs signifies a crucial step towards advancing communication technologies beyond Earth, opening new frontiers for space exploration and scientific discovery.

\subsection{Quantum-assisted Digital Twins for NTNs}
Quantum-assisted digital twins represent an innovative approach aim to further develop efficient NTNs by leveraging the capabilities of quantum technologies, including sensing, computing, and security. 
A quantum-based digital twin is a virtual representation of a NTN that provides real-time insights into the network's performance and behavior \cite{al2023digital}.
In this paradigm, quantum sensing technologies enhance the accuracy and precision of data collection, enabling the creation of highly detailed digital replicas of physical entities in space. Quantum computing plays a pivotal role in processing the vast amounts of data generated by these digital twins, performing complex simulations and analyses with unprecedented speed and efficiency. Moreover, the inherent security features of quantum technologies ensure the integrity and confidentiality of the digital twin data, safeguarding critical information in NTNs. This holistic integration of quantum technologies into digital twins for NTNs promises enhanced performance, reliability, and security, ultimately advancing the capabilities of space-based networks for diverse applications.




\subsection{Quantum Communications for Healthcare}
One of the sensitive issues in digital healthcare is encryption and security of data. Quantum communications provides methods of secure exchange of health records by QKD and anonymous private information retrieval systems \cite{KKu:22:TCOM}. {Additionally, security of medical media is imperative for patient safety and confidentiality, and thus, recently the concept of quantum medical image encryption has attracted a significant attention from both scientists and healthcare system designers\cite{Abd2018}. In this framework, medical images and records can be securely communicated within different health centers using quantum encryption/decryption algorithms.}
Another interesting feature is to offer certified deletion of health records that generates a classical certificate of deletion of health records \cite{HMN:21:AC}. These features make quantum communications systems attractive for digital healthcare solutions and other databases of sensitive nature. 
{Furthermore, quantum computing can also help in this context via optimizing the healthcare system models to advance the patient care experience, improve the population health, and minimize per capita healthcare costs \cite{jayanthi2022}.}

\subsection{Quantum for Banking and Finance Industries}
Banking and finance industry have strict requirements for encryption due to sensitive nature of their operations and data. On the one hand, banks and financial institutions require real-time encryption capabilities for the large-volume of their real-time transactions, which is a major growing challenge. Introducing quantum to NTNs offers a solution to this challenge in the form of satellite-based QKD with the possibility of global connectivity. On the other hand, quantum computing also offers appealing solutions for the finance sector in the form of quantum algorithms for risk-based asset management, portfolio optimization, and other complicated financial procedures  \cite{HBH:19:arXiv, CKA:21:arXiv}. Specifically, quantum computing can further develop the investment industry via applying quantum-based machine learning algorithms for managing massive numbers of underlying assets while considering various sets of relevant data for learning, adapting, and enhancing investment decisions. Beyond this, with the availability of cloud quantum computers and the possibility of blind quantum computation, there exists an opportunity to put these quantum solutions to test and harness their benefits \cite{Fitzsimons2017}. Interested readers may refer to the recent article in \cite{Herman2023}, which provides a comprehensive summary of the current state of quantum computing for financial applications, particularly emphasizing stochastic modeling, optimization, and machine learning.

\subsection{Quantum Technologies for Government and Defense}
Communication within the governmental organizations and defense establishments are under persistent threats of espionage and cyber-attacks. The unconditional security offered by the QKD and other quantum encryption techniques is an effective countermeasure to protect against these threats. Furthermore, quantum technologies including communications, computing, and sensing are offering a set of beneficial tools and mechanisms for defense and military applications \cite{Krelina2021}.
For instance, quantum sensors can be used to detect submarines and stealth aircraft \cite{Gamberini2021}. Specifically, utilizing quantum sensors for positioning, navigation and timing can induce reliable inertial navigation systems, which empower navigation without the need for external references. A gravimeter based on quantum sensing has been proposed in \cite{Johnsson2016} to detect changes in the gravitational field. This gravimeter uses a quantum magnetomechanical system consisting of a magnetically trapped superconducting resonator, and it is a passive system that probes without transmitting signals. This allows the detection of objects, which may not emit any kind of electromagnetic signals, by only observing the surrounding transient gravitational changes.

\section{Conclusions}\label{sec:conclusions} 
Our exploration into the interplay between quantum technologies and NTNs revealed a promising landscape of opportunities and advancements. Through a comprehensive review of the integration potential, we have identified key synergies that can significantly enhance the capabilities of NTNs and expand the reach of quantum communication. The unique attributes of quantum technologies, such as unconditional security, large communication capacity, computational speed, and precise sensing, can provide a foundation for groundbreaking applications within NTNs. In addressing the challenges associated with quantum-NTN integration, we have discussed potential solutions to overcome the complexities arising from dynamic propagation environments, scalability concerns, network challenges, and resource management intricacies. Moreover, we have highlighted various innovative visions and research directions motivated by the utilization of quantum technologies in non-terrestrial communication systems. As we navigate toward the next-generation of wireless networks, harnessing the power of quantum technologies within NTNs emerges as a strategic imperative to achieve unprecedented levels of reliability, efficiency, and security in future communication systems. In short, this article explores the integration of quantum technologies into NTNs, with the aim of inspiring in-depth investigations and motivating further research endeavors in these domains.


\linespread{1.13}

\bibliographystyle{IEEEtran}
\bibliography{IEEEabrv,References}

\end{document}

%% file: Paper_structure.pdf_tex
\begingroup%
  \makeatletter%
  \providecommand\color[2][]{%
    \errmessage{(Inkscape) Color is used for the text in Inkscape, but the package 'color.sty' is not loaded}%
    \renewcommand\color[2][]{}%
  }%
  \providecommand\transparent[1]{%
    \errmessage{(Inkscape) Transparency is used (non-zero) for the text in Inkscape, but the package 'transparent.sty' is not loaded}%
    \renewcommand\transparent[1]{}%
  }%
  \providecommand\rotatebox[2]{#2}%
  \newcommand*\fsize{\dimexpr\f@size pt\relax}%
  \newcommand*\lineheight[1]{\fontsize{\fsize}{#1\fsize}\selectfont}%
  \ifx\svgwidth\undefined%
    \setlength{\unitlength}{568.77449096bp}%
    \ifx\svgscale\undefined%
      \relax%
    \else%
      \setlength{\unitlength}{\unitlength * \real{\svgscale}}%
    \fi%
  \else%
    \setlength{\unitlength}{\svgwidth}%
  \fi%
  \global\let\svgwidth\undefined%
  \global\let\svgscale\undefined%
  \makeatother%
  \begin{picture}(1,0.8813387)%
    \lineheight{1}%
    \setlength\tabcolsep{0pt}%
    \put(0,0){\includegraphics[width=\unitlength,page=1]{Paper_structure.pdf}}%
    \put(0.27511323,0.85371383){\color[rgb]{0,0,0}\makebox(0,0)[lt]{\lineheight{1.25}\smash{\begin{tabular}[t]{l}\textbf{Section I}\end{tabular}}}}%
    \put(0.26462012,0.82991383){\color[rgb]{0,0,0}\makebox(0,0)[lt]{\lineheight{1.25}\smash{\begin{tabular}[t]{l}Introduction\end{tabular}}}}%
    \put(0,0){\includegraphics[width=\unitlength,page=2]{Paper_structure.pdf}}%
    \put(0.26960297,0.54653009){\color[rgb]{0,0,0}\makebox(0,0)[lt]{\lineheight{1.25}\smash{\begin{tabular}[t]{l}\textbf{Section III}\end{tabular}}}}%
    \put(0.20072362,0.52273003){\color[rgb]{0,0,0}\makebox(0,0)[lt]{\lineheight{1.25}\smash{\begin{tabular}[t]{l}  \end{tabular}}}}%
    \put(0.20198765,0.52009278){\color[rgb]{0,0,0}\makebox(0,0)[lt]{\lineheight{1.25}\smash{\begin{tabular}[t]{l}Non-Terrestrial Networks\end{tabular}}}}%
    \put(0,0){\includegraphics[width=\unitlength,page=3]{Paper_structure.pdf}}%
    \put(0.32221701,0.73848243){\color[rgb]{0,0,0}\makebox(0,0)[t]{\lineheight{1.25}\smash{\begin{tabular}[t]{c}\textbf{Section II}\\Quantum Technologies\end{tabular}}}}%
    \put(0,0){\includegraphics[width=\unitlength,page=4]{Paper_structure.pdf}}%
    \put(0.31963824,0.36442094){\color[rgb]{0,0,0}\makebox(0,0)[t]{\lineheight{1.25}\smash{\begin{tabular}[t]{c}\textbf{Section IV}\\Integration Challenges and Solutions\end{tabular}}}}%
    \put(0.14722267,0.33798378){\color[rgb]{0,0,0}\makebox(0,0)[lt]{\lineheight{1.25}\smash{\begin{tabular}[t]{l} \end{tabular}}}}%
    \put(0,0){\includegraphics[width=\unitlength,page=5]{Paper_structure.pdf}}%
    \put(0.26849914,0.03823863){\color[rgb]{0,0,0}\makebox(0,0)[lt]{\lineheight{1.25}\smash{\begin{tabular}[t]{l}\textbf{Section VI}\end{tabular}}}}%
    \put(0.26242187,0.01443861){\color[rgb]{0,0,0}\makebox(0,0)[lt]{\lineheight{1.25}\smash{\begin{tabular}[t]{l}Conclusions\end{tabular}}}}%
    \put(0,0){\includegraphics[width=\unitlength,page=6]{Paper_structure.pdf}}%
    \put(0.27125403,0.16245779){\color[rgb]{0,0,0}\makebox(0,0)[lt]{\lineheight{1.25}\smash{\begin{tabular}[t]{l}\textbf{Section V}\end{tabular}}}}%
    \put(0.16540989,0.13865777){\color[rgb]{0,0,0}\makebox(0,0)[lt]{\lineheight{1.25}\smash{\begin{tabular}[t]{l}Future Directions and Applications\end{tabular}}}}%
    \put(0,0){\includegraphics[width=\unitlength,page=7]{Paper_structure.pdf}}%
    \put(0.5536445,0.79798231){\color[rgb]{0,0,0}\makebox(0,0)[lt]{\lineheight{1.25}\smash{\begin{tabular}[t]{l}  \end{tabular}}}}%
    \put(0.56466499,0.7821588){\color[rgb]{0,0,0}\makebox(0,0)[lt]{\lineheight{1.25}\smash{\begin{tabular}[t]{l}A. Quantum Information: Basic Theory\\B. Quantum Communications\\C. Quantum Computing\\D. Quantum Sensing\\E. Quantum Key Distribution (QKD)\\\end{tabular}}}}%
    \put(0.5536445,0.77418231){\color[rgb]{0,0,0}\makebox(0,0)[lt]{\lineheight{1.25}\smash{\begin{tabular}[t]{l}  \end{tabular}}}}%
    \put(0.5536445,0.75038231){\color[rgb]{0,0,0}\makebox(0,0)[lt]{\lineheight{1.25}\smash{\begin{tabular}[t]{l}  \end{tabular}}}}%
    \put(0.5536445,0.72658231){\color[rgb]{0,0,0}\makebox(0,0)[lt]{\lineheight{1.25}\smash{\begin{tabular}[t]{l}       \end{tabular}}}}%
    \put(0.5536445,0.70278235){\color[rgb]{0,0,0}\makebox(0,0)[lt]{\lineheight{1.25}\smash{\begin{tabular}[t]{l}       \end{tabular}}}}%
    \put(0.5536445,0.67898233){\color[rgb]{0,0,0}\makebox(0,0)[lt]{\lineheight{1.25}\smash{\begin{tabular}[t]{l}       \end{tabular}}}}%
    \put(0.5536445,0.65518235){\color[rgb]{0,0,0}\makebox(0,0)[lt]{\lineheight{1.25}\smash{\begin{tabular}[t]{l}       \end{tabular}}}}%
    \put(0,0){\includegraphics[width=\unitlength,page=8]{Paper_structure.pdf}}%
    \put(0.03526108,0.38213839){\color[rgb]{0,0,0}\rotatebox{90}{\makebox(0,0)[lt]{\lineheight{1.25}\smash{\begin{tabular}[t]{l}\textbf{Paper structure}\end{tabular}}}}}%
    \put(0,0){\includegraphics[width=\unitlength,page=9]{Paper_structure.pdf}}%
    \put(0.5536445,0.58223011){\color[rgb]{0,0,0}\makebox(0,0)[lt]{\lineheight{1.25}\smash{\begin{tabular}[t]{l}  \end{tabular}}}}%
    \put(0.56466511,0.57168111){\color[rgb]{0,0,0}\makebox(0,0)[lt]{\lineheight{1.25}\smash{\begin{tabular}[t]{l}A. General Description\\B. Quantum-NTN Synergies\\\end{tabular}}}}%
    \put(0.5536445,0.5584301){\color[rgb]{0,0,0}\makebox(0,0)[lt]{\lineheight{1.25}\smash{\begin{tabular}[t]{l}  \end{tabular}}}}%
    \put(0.5536445,0.53463012){\color[rgb]{0,0,0}\makebox(0,0)[lt]{\lineheight{1.25}\smash{\begin{tabular}[t]{l}      \end{tabular}}}}%
    \put(0.5536445,0.51083014){\color[rgb]{0,0,0}\makebox(0,0)[lt]{\lineheight{1.25}\smash{\begin{tabular}[t]{l}       \end{tabular}}}}%
    \put(0.5536445,0.48703016){\color[rgb]{0,0,0}\makebox(0,0)[lt]{\lineheight{1.25}\smash{\begin{tabular}[t]{l}       \end{tabular}}}}%
    \put(0,0){\includegraphics[width=\unitlength,page=10]{Paper_structure.pdf}}%
    \put(0.5536445,0.40938361){\color[rgb]{0,0,0}\makebox(0,0)[lt]{\lineheight{1.25}\smash{\begin{tabular}[t]{l}  \end{tabular}}}}%
    \put(0.56466511,0.40938361){\color[rgb]{0,0,0}\makebox(0,0)[lt]{\lineheight{1.25}\smash{\begin{tabular}[t]{l}A. Channel Reliability\end{tabular}}}}%
    \put(0.5536445,0.38558374){\color[rgb]{0,0,0}\makebox(0,0)[lt]{\lineheight{1.25}\smash{\begin{tabular}[t]{l}  \end{tabular}}}}%
    \put(0.56466511,0.38558374){\color[rgb]{0,0,0}\makebox(0,0)[lt]{\lineheight{1.25}\smash{\begin{tabular}[t]{l}B. Network Flexibility and Reconfigurability\end{tabular}}}}%
    \put(0.5536445,0.36178373){\color[rgb]{0,0,0}\makebox(0,0)[lt]{\lineheight{1.25}\smash{\begin{tabular}[t]{l}  \end{tabular}}}}%
    \put(0.56466511,0.36178373){\color[rgb]{0,0,0}\makebox(0,0)[lt]{\lineheight{1.25}\smash{\begin{tabular}[t]{l}C. Network Scalability\end{tabular}}}}%
    \put(0.5536445,0.33798371){\color[rgb]{0,0,0}\makebox(0,0)[lt]{\lineheight{1.25}\smash{\begin{tabular}[t]{l}       \end{tabular}}}}%
    \put(0.58957905,0.33798371){\color[rgb]{0,0,0}\makebox(0,0)[lt]{\lineheight{1.25}\smash{\begin{tabular}[t]{l}1) Trusted Relays\end{tabular}}}}%
    \put(0.5536445,0.31418369){\color[rgb]{0,0,0}\makebox(0,0)[lt]{\lineheight{1.25}\smash{\begin{tabular}[t]{l}       \end{tabular}}}}%
    \put(0.58957905,0.31418369){\color[rgb]{0,0,0}\makebox(0,0)[lt]{\lineheight{1.25}\smash{\begin{tabular}[t]{l}2) Untrusted Measurement Nodes\end{tabular}}}}%
    \put(0.5536445,0.29038367){\color[rgb]{0,0,0}\makebox(0,0)[lt]{\lineheight{1.25}\smash{\begin{tabular}[t]{l}       \end{tabular}}}}%
    \put(0.58957905,0.29038367){\color[rgb]{0,0,0}\makebox(0,0)[lt]{\lineheight{1.25}\smash{\begin{tabular}[t]{l}3) Quantum Repeaters\end{tabular}}}}%
    \put(0,0){\includegraphics[width=\unitlength,page=11]{Paper_structure.pdf}}%
    \put(0.5536445,0.21005767){\color[rgb]{0,0,0}\makebox(0,0)[lt]{\lineheight{1.25}\smash{\begin{tabular}[t]{l}  \end{tabular}}}}%
    \put(0.5536445,0.18625773){\color[rgb]{0,0,0}\makebox(0,0)[lt]{\lineheight{1.25}\smash{\begin{tabular}[t]{l}  \end{tabular}}}}%
    \put(0.5536445,0.16245779){\color[rgb]{0,0,0}\makebox(0,0)[lt]{\lineheight{1.25}\smash{\begin{tabular}[t]{l}  \end{tabular}}}}%
    \put(0.5536445,0.13865777){\color[rgb]{0,0,0}\makebox(0,0)[lt]{\lineheight{1.25}\smash{\begin{tabular}[t]{l}  \end{tabular}}}}%
    \put(0.5536445,0.11485775){\color[rgb]{0,0,0}\makebox(0,0)[lt]{\lineheight{1.25}\smash{\begin{tabular}[t]{l}  \end{tabular}}}}%
    \put(0.5536445,0.09105773){\color[rgb]{0,0,0}\makebox(0,0)[lt]{\lineheight{1.25}\smash{\begin{tabular}[t]{l}      \end{tabular}}}}%
    \put(0,0){\includegraphics[width=\unitlength,page=12]{Paper_structure.pdf}}%
    \put(0.56466499,0.22570164){\color[rgb]{0,0,0}\makebox(0,0)[lt]{\lineheight{1.25}\smash{\begin{tabular}[t]{l}A. Space-based Quantum Clouds\\B. Quantum Computing for Space Missions\\C. Deep Space Quantum Communication\\D. Quantum-assisted Digital Twins for NTNs\\E. Quantum Communications for Healthcare\\F. Quantum for Banking and Finance Industries\\G. Quantum for Government and Defense\\\end{tabular}}}}%
    \put(0.58957905,0.52258999){\color[rgb]{0,0,0}\makebox(0,0)[lt]{\lineheight{1.25}\smash{\begin{tabular}[t]{l}1) Quantum over NTN Links\\2) Beyond-QKD Applications\\\end{tabular}}}}%
  \end{picture}%
\endgroup%

%% file: multilayer_quantum_NTN.pdf_tex
\begingroup%
  \makeatletter%
  \providecommand\color[2][]{%
    \errmessage{(Inkscape) Color is used for the text in Inkscape, but the package 'color.sty' is not loaded}%
    \renewcommand\color[2][]{}%
  }%
  \providecommand\transparent[1]{%
    \errmessage{(Inkscape) Transparency is used (non-zero) for the text in Inkscape, but the package 'transparent.sty' is not loaded}%
    \renewcommand\transparent[1]{}%
  }%
  \providecommand\rotatebox[2]{#2}%
  \newcommand*\fsize{\dimexpr\f@size pt\relax}%
  \newcommand*\lineheight[1]{\fontsize{\fsize}{#1\fsize}\selectfont}%
  \ifx\svgwidth\undefined%
    \setlength{\unitlength}{305.82103098bp}%
    \ifx\svgscale\undefined%
      \relax%
    \else%
      \setlength{\unitlength}{\unitlength * \real{\svgscale}}%
    \fi%
  \else%
    \setlength{\unitlength}{\svgwidth}%
  \fi%
  \global\let\svgwidth\undefined%
  \global\let\svgscale\undefined%
  \makeatother%
  \begin{picture}(1,0.96633564)%
    \lineheight{1}%
    \setlength\tabcolsep{0pt}%
    \put(0,0){\includegraphics[width=\unitlength,page=1]{multilayer_quantum_NTN.pdf}}%
    \put(0.1332692,0.14080145){\color[rgb]{0,0,0}\makebox(0,0)[lt]{\lineheight{1.25}\smash{\begin{tabular}[t]{l}ISL\\FSO\\ECL\end{tabular}}}}%
    \put(0,0){\includegraphics[width=\unitlength,page=2]{multilayer_quantum_NTN.pdf}}%
    \put(0.90413553,0.66371353){\color[rgb]{0,0,0}\makebox(0,0)[t]{\lineheight{1.25}\smash{\begin{tabular}[t]{c}NGSO\\satellites\end{tabular}}}}%
    \put(0.90413553,0.91371069){\color[rgb]{0,0,0}\makebox(0,0)[t]{\lineheight{1.25}\smash{\begin{tabular}[t]{c}GSO\\satellites\end{tabular}}}}%
    \put(0.90832181,0.38168553){\color[rgb]{0,0,0}\makebox(0,0)[t]{\lineheight{1.25}\smash{\begin{tabular}[t]{c}Aerial\\platforms\end{tabular}}}}%
    \put(0.90743093,0.09965817){\color[rgb]{0,0,0}\makebox(0,0)[t]{\lineheight{1.25}\smash{\begin{tabular}[t]{c}Ground\\network\end{tabular}}}}%
    \put(0,0){\includegraphics[width=\unitlength,page=3]{multilayer_quantum_NTN.pdf}}%
  \end{picture}%
\endgroup%

%% file: Quantum_ch.pdf_tex
\begingroup%
  \makeatletter%
  \providecommand\color[2][]{%
    \errmessage{(Inkscape) Color is used for the text in Inkscape, but the package 'color.sty' is not loaded}%
    \renewcommand\color[2][]{}%
  }%
  \providecommand\transparent[1]{%
    \errmessage{(Inkscape) Transparency is used (non-zero) for the text in Inkscape, but the package 'transparent.sty' is not loaded}%
    \renewcommand\transparent[1]{}%
  }%
  \providecommand\rotatebox[2]{#2}%
  \newcommand*\fsize{\dimexpr\f@size pt\relax}%
  \newcommand*\lineheight[1]{\fontsize{\fsize}{#1\fsize}\selectfont}%
  \ifx\svgwidth\undefined%
    \setlength{\unitlength}{585.70185992bp}%
    \ifx\svgscale\undefined%
      \relax%
    \else%
      \setlength{\unitlength}{\unitlength * \real{\svgscale}}%
    \fi%
  \else%
    \setlength{\unitlength}{\svgwidth}%
  \fi%
  \global\let\svgwidth\undefined%
  \global\let\svgscale\undefined%
  \makeatother%
  \begin{picture}(1,0.20482641)%
    \lineheight{1}%
    \setlength\tabcolsep{0pt}%
    \put(0,0){\includegraphics[width=\unitlength,page=1]{Quantum_ch.pdf}}%
    \put(0.06732545,0.159463){\color[rgb]{0,0,0}\makebox(0,0)[t]{\lineheight{1.11000001}\smash{\begin{tabular}[t]{c}Classical\\Information\end{tabular}}}}%
    \put(0.25288148,0.159463){\color[rgb]{0,0,0}\makebox(0,0)[t]{\lineheight{1.10000002}\smash{\begin{tabular}[t]{c}Quantum\\States\end{tabular}}}}%
    \put(0,0){\includegraphics[width=\unitlength,page=2]{Quantum_ch.pdf}}%
    \put(0.49481306,0.17313134){\color[rgb]{0,0,0}\makebox(0,0)[t]{\lineheight{1.10000002}\smash{\begin{tabular}[t]{c}\textbf{NTN Quantum Channel}\\Free space optical links or\\Li-Fi links \end{tabular}}}}%
    \put(0.91374367,0.159463){\color[rgb]{0,0,0}\makebox(0,0)[t]{\lineheight{1.10000002}\smash{\begin{tabular}[t]{c}Classical\\Information\end{tabular}}}}%
    \put(0.73435738,0.159463){\color[rgb]{0,0,0}\makebox(0,0)[t]{\lineheight{1.10000002}\smash{\begin{tabular}[t]{c}Quantum\\States\end{tabular}}}}%
    \put(0,0){\includegraphics[width=\unitlength,page=3]{Quantum_ch.pdf}}%
    \put(0.02592359,0.01294243){\color[rgb]{0,0,0}\makebox(0,0)[lt]{\lineheight{1.25}\smash{\begin{tabular}[t]{l} \end{tabular}}}}%
    \put(0.18767574,0.04668866){\color[rgb]{0,0,0}\makebox(0,0)[t]{\smash{\begin{tabular}[t]{c}\textbf{Transmitter side}\\Preparation (encoding) classical information\\ into quantum states\end{tabular}}}}%
    \put(0.82910454,0.04458176){\color[rgb]{0,0,0}\makebox(0,0)[t]{\smash{\begin{tabular}[t]{c}\textbf{Receiver side}\\Measurement (decoding) quantum states \\to obtain classical information\end{tabular}}}}%
    \put(0,0){\includegraphics[width=\unitlength,page=4]{Quantum_ch.pdf}}%
  \end{picture}%
\endgroup%

%% file: MPE2.pdf_tex
\begingroup%
  \makeatletter%
  \providecommand\color[2][]{%
    \errmessage{(Inkscape) Color is used for the text in Inkscape, but the package 'color.sty' is not loaded}%
    \renewcommand\color[2][]{}%
  }%
  \providecommand\transparent[1]{%
    \errmessage{(Inkscape) Transparency is used (non-zero) for the text in Inkscape, but the package 'transparent.sty' is not loaded}%
    \renewcommand\transparent[1]{}%
  }%
  \providecommand\rotatebox[2]{#2}%
  \newcommand*\fsize{\dimexpr\f@size pt\relax}%
  \newcommand*\lineheight[1]{\fontsize{\fsize}{#1\fsize}\selectfont}%
  \ifx\svgwidth\undefined%
    \setlength{\unitlength}{382.81479723bp}%
    \ifx\svgscale\undefined%
      \relax%
    \else%
      \setlength{\unitlength}{\unitlength * \real{\svgscale}}%
    \fi%
  \else%
    \setlength{\unitlength}{\svgwidth}%
  \fi%
  \global\let\svgwidth\undefined%
  \global\let\svgscale\undefined%
  \makeatother%
  \begin{picture}(1,1.22509797)%
    \lineheight{1}%
    \setlength\tabcolsep{0pt}%
    \put(0.43436148,1.0033567){\color[rgb]{0,0,0}\makebox(0,0)[lt]{\lineheight{1.25}\smash{\begin{tabular}[t]{l}\large$q_0$\end{tabular}}}}%
    \put(0.90824243,0.82259398){\color[rgb]{0,0,0}\makebox(0,0)[lt]{\lineheight{1.25}\smash{\begin{tabular}[t]{l}\large$q_1$\end{tabular}}}}%
    \put(0.99220667,0.40780967){\color[rgb]{0,0,0}\makebox(0,0)[lt]{\lineheight{1.25}\smash{\begin{tabular}[t]{l}\large$q_2$\end{tabular}}}}%
    \put(0.73219677,0.08678549){\color[rgb]{0,0,0}\makebox(0,0)[lt]{\lineheight{1.25}\smash{\begin{tabular}[t]{l}\large$q_3$\end{tabular}}}}%
    \put(-0.04489471,0.43425833){\color[rgb]{0,0,0}\makebox(0,0)[lt]{\lineheight{1.25}\smash{\begin{tabular}[t]{l}\large$q_5$\end{tabular}}}}%
    \put(0.06481897,0.86709499){\color[rgb]{0,0,0}\makebox(0,0)[lt]{\lineheight{1.25}\smash{\begin{tabular}[t]{l}\large$q_{N-1}$\end{tabular}}}}%
    \put(0.51844032,0.57106216){\color[rgb]{0,0,0}\makebox(0,0)[t]{\lineheight{1.5}\smash{\begin{tabular}[t]{c}Multipartite Entanglement for: \\Quantum Conference Key Agreement \\Anonymous Entanglement Generation \\Quantum Anonymous Network \\Quantum Secret Sharing \end{tabular}}}}%
    \put(0,0){\includegraphics[width=\unitlength,page=1]{MPE2.pdf}}%
    \put(0.49013386,0.69320564){\color[rgb]{0,0,0}\makebox(0,0)[t]{\lineheight{1.5}\smash{\begin{tabular}[t]{c}\large{$\ket{\psi}_{q_0 \cdots q_{N-1}}=\frac{\ket{0}^{\otimes N}+ \ket{1}^{\otimes N} }{\sqrt{2}}$}\end{tabular}}}}%
    \put(0,0){\includegraphics[width=\unitlength,page=2]{MPE2.pdf}}%
    \put(0.44522133,1.07995167){\color[rgb]{0,0,0}\makebox(0,0)[lt]{\lineheight{1.25}\smash{\begin{tabular}[t]{l}TP\end{tabular}}}}%
    \put(0,0){\includegraphics[width=\unitlength,page=3]{MPE2.pdf}}%
    \put(0.2060664,0.09895841){\color[rgb]{0,0,0}\makebox(0,0)[lt]{\lineheight{1.25}\smash{\begin{tabular}[t]{l}\large$q_4$\end{tabular}}}}%
    \put(0,0){\includegraphics[width=\unitlength,page=4]{MPE2.pdf}}%
  \end{picture}%
\endgroup%

%% file: QMIMO_Channel.pdf_tex
\begingroup%
  \makeatletter%
  \providecommand\color[2][]{%
    \errmessage{(Inkscape) Color is used for the text in Inkscape, but the package 'color.sty' is not loaded}%
    \renewcommand\color[2][]{}%
  }%
  \providecommand\transparent[1]{%
    \errmessage{(Inkscape) Transparency is used (non-zero) for the text in Inkscape, but the package 'transparent.sty' is not loaded}%
    \renewcommand\transparent[1]{}%
  }%
  \providecommand\rotatebox[2]{#2}%
  \newcommand*\fsize{\dimexpr\f@size pt\relax}%
  \newcommand*\lineheight[1]{\fontsize{\fsize}{#1\fsize}\selectfont}%
  \ifx\svgwidth\undefined%
    \setlength{\unitlength}{435.27272469bp}%
    \ifx\svgscale\undefined%
      \relax%
    \else%
      \setlength{\unitlength}{\unitlength * \real{\svgscale}}%
    \fi%
  \else%
    \setlength{\unitlength}{\svgwidth}%
  \fi%
  \global\let\svgwidth\undefined%
  \global\let\svgscale\undefined%
  \makeatother%
  \begin{picture}(1,0.35295457)%
    \lineheight{1}%
    \setlength\tabcolsep{0pt}%
    \put(0,0){\includegraphics[width=\unitlength,page=1]{QMIMO_Channel.pdf}}%
    \put(0.19179818,0.21188842){\color[rgb]{0,0,0}\makebox(0,0)[t]{\lineheight{0.89999998}\smash{\begin{tabular}[t]{c}Quantum\\Encoder\end{tabular}}}}%
    \put(0,0){\includegraphics[width=\unitlength,page=2]{QMIMO_Channel.pdf}}%
    \put(0.78291638,0.22656922){\color[rgb]{0,0,0}\makebox(0,0)[t]{\smash{\begin{tabular}[t]{c}Quantum\\Measuring \\Apparatus\end{tabular}}}}%
    \put(0.51901976,0.00554562){\color[rgb]{0,0,0}\makebox(0,0)[t]{\lineheight{1.25}\smash{\begin{tabular}[t]{c}Quantum MIMO Channel\end{tabular}}}}%
    \put(0,0){\includegraphics[width=\unitlength,page=3]{QMIMO_Channel.pdf}}%
    \put(0.38463206,0.1795169){\color[rgb]{0,0,0}\makebox(0,0)[lt]{\lineheight{1.25}\smash{\begin{tabular}[t]{l}\textbf{.}\end{tabular}}}}%
    \put(0.38463206,0.16538105){\color[rgb]{0,0,0}\makebox(0,0)[lt]{\lineheight{1.25}\smash{\begin{tabular}[t]{l}\textbf{.}\end{tabular}}}}%
    \put(0.38463206,0.15124441){\color[rgb]{0,0,0}\makebox(0,0)[lt]{\lineheight{1.25}\smash{\begin{tabular}[t]{l}\textbf{.}\end{tabular}}}}%
    \put(0,0){\includegraphics[width=\unitlength,page=4]{QMIMO_Channel.pdf}}%
    \put(0.61347421,0.30202182){\color[rgb]{0,0,0}\makebox(0,0)[lt]{\lineheight{1.25}\smash{\begin{tabular}[t]{l}\textit{$O_1$}\end{tabular}}}}%
    \put(0,0){\includegraphics[width=\unitlength,page=5]{QMIMO_Channel.pdf}}%
    \put(0.61347421,0.24366392){\color[rgb]{0,0,0}\makebox(0,0)[lt]{\lineheight{1.25}\smash{\begin{tabular}[t]{l}\textit{$O_2$}\end{tabular}}}}%
    \put(0,0){\includegraphics[width=\unitlength,page=6]{QMIMO_Channel.pdf}}%
    \put(0.61538604,0.0788397){\color[rgb]{0,0,0}\makebox(0,0)[lt]{\lineheight{1.25}\smash{\begin{tabular}[t]{l}\textit{$O_N$}\end{tabular}}}}%
    \put(0.62462243,0.18032399){\color[rgb]{0,0,0}\makebox(0,0)[lt]{\lineheight{1.25}\smash{\begin{tabular}[t]{l}\textbf{.}\end{tabular}}}}%
    \put(0.62462243,0.16618794){\color[rgb]{0,0,0}\makebox(0,0)[lt]{\lineheight{1.25}\smash{\begin{tabular}[t]{l}\textbf{.}\end{tabular}}}}%
    \put(0.62462243,0.15205149){\color[rgb]{0,0,0}\makebox(0,0)[lt]{\lineheight{1.25}\smash{\begin{tabular}[t]{l}\textbf{.}\end{tabular}}}}%
    \put(0,0){\includegraphics[width=\unitlength,page=7]{QMIMO_Channel.pdf}}%
    \put(0.92128264,0.18965766){\color[rgb]{0,0,0}\makebox(0,0)[lt]{\lineheight{1.25}\smash{\begin{tabular}[t]{l}Output\end{tabular}}}}%
    \put(0,0){\includegraphics[width=\unitlength,page=8]{QMIMO_Channel.pdf}}%
    \put(-0.0007053,0.19542611){\color[rgb]{0,0,0}\makebox(0,0)[lt]{\lineheight{1.25}\smash{\begin{tabular}[t]{l}Input\end{tabular}}}}%
    \put(0.40110982,0.30330291){\color[rgb]{0,0,0}\makebox(0,0)[lt]{\lineheight{1.25}\smash{\begin{tabular}[t]{l}\textit{$I_1$}\end{tabular}}}}%
    \put(0.40110982,0.24471805){\color[rgb]{0,0,0}\makebox(0,0)[lt]{\lineheight{1.25}\smash{\begin{tabular}[t]{l}\textit{$I_2$}\end{tabular}}}}%
    \put(0.40110982,0.07930197){\color[rgb]{0,0,0}\makebox(0,0)[lt]{\lineheight{1.25}\smash{\begin{tabular}[t]{l}\textit{$I_N$}\end{tabular}}}}%
    \put(0,0){\includegraphics[width=\unitlength,page=9]{QMIMO_Channel.pdf}}%
  \end{picture}%
\endgroup%

%% file: Sat_QKD.pdf_tex
\begingroup%
  \makeatletter%
  \providecommand\color[2][]{%
    \errmessage{(Inkscape) Color is used for the text in Inkscape, but the package 'color.sty' is not loaded}%
    \renewcommand\color[2][]{}%
  }%
  \providecommand\transparent[1]{%
    \errmessage{(Inkscape) Transparency is used (non-zero) for the text in Inkscape, but the package 'transparent.sty' is not loaded}%
    \renewcommand\transparent[1]{}%
  }%
  \providecommand\rotatebox[2]{#2}%
  \newcommand*\fsize{\dimexpr\f@size pt\relax}%
  \newcommand*\lineheight[1]{\fontsize{\fsize}{#1\fsize}\selectfont}%
  \ifx\svgwidth\undefined%
    \setlength{\unitlength}{422.77423528bp}%
    \ifx\svgscale\undefined%
      \relax%
    \else%
      \setlength{\unitlength}{\unitlength * \real{\svgscale}}%
    \fi%
  \else%
    \setlength{\unitlength}{\svgwidth}%
  \fi%
  \global\let\svgwidth\undefined%
  \global\let\svgscale\undefined%
  \makeatother%
  \begin{picture}(1,0.69566909)%
    \lineheight{1}%
    \setlength\tabcolsep{0pt}%
    \put(0,0){\includegraphics[width=\unitlength,page=1]{Sat_QKD.pdf}}%
    \put(0.18082712,0.44297966){\color[rgb]{0,0,0}\makebox(0,0)[lt]{\lineheight{1.20000005}\smash{\begin{tabular}[t]{l}$t=0 : u_1 \longleftrightarrow u_2$ secret link required\\$t=1 : u_1$ sends request to $s_1$\\$t=2 : s_1$ discovers $s_2$ in range of $u_2$\\$t=3 : k_1=$QKD$(u_1,s_1)$\\           $\hspace{9mm}k_2=$QKD$(s_1,s_2)$\\           $\hspace{9mm}k_2=$QKD$(s_2,u_2)$\\$t=4 : s_1 \longrightarrow s_2$ Classical message $k_1\otimes k_2$\\$t=5 : s_2 \longrightarrow u_2$ Classical message $k_1\otimes k_3$\\$t=6 :$ ACK sent,  $u_2$ and  $u_2$ share $k_1$\\\end{tabular}}}}%
    \put(0.06338739,-0.02738233){\color[rgb]{0,0,0}\makebox(0,0)[t]{\lineheight{1.5}\smash{\begin{tabular}[t]{c}Transmitter: $u_1$\end{tabular}}}}%
    \put(0.94328897,-0.02738223){\color[rgb]{0,0,0}\makebox(0,0)[t]{\lineheight{1.5}\smash{\begin{tabular}[t]{c}Receiver: $u_2$\end{tabular}}}}%
    \put(0,0){\includegraphics[width=\unitlength,page=2]{Sat_QKD.pdf}}%
    \put(0.13617712,0.69783594){\color[rgb]{0,0,0}\makebox(0,0)[t]{\lineheight{1.5}\smash{\begin{tabular}[t]{c}Trusted Relay: $s_1$\end{tabular}}}}%
    \put(0.8533785,0.70163738){\color[rgb]{0,0,0}\makebox(0,0)[t]{\lineheight{1.5}\smash{\begin{tabular}[t]{c}Trusted Relay: $s_2$\end{tabular}}}}%
    \put(0,0){\includegraphics[width=\unitlength,page=3]{Sat_QKD.pdf}}%
  \end{picture}%
\endgroup%